\documentclass[a4paper,11pt]{article}
\pdfoutput=1
\usepackage{jheppub}
\usepackage[T1]{fontenc} % if needed
\usepackage{amsmath}
\usepackage{amsfonts}
\usepackage{amssymb}
\usepackage{mathrsfs}
\usepackage{mathtools}
\usepackage[english]{babel}

\title{\boldmath Axion Inflation with an SU(2) Gauge Field: Detectable Chiral Gravity Waves }

\author{Azadeh Maleknejad}

\affiliation{School of Physics, Institute for Research in Fundamental Sciences (IPM),\\ P. Code. 19538-33511, Tehran, Iran}
\emailAdd{azade@ipm.ir}

\abstract{We study a single field axion inflation model in the presence of an SU(2) gauge field with a small vev.
In order to make the analysis as model-independent as possible, we consider an arbitrary potential for the axion that is able to support the slow-roll inflation. The gauge field is coupled to the axion with a Chern-Simons interaction $\frac{\lambda}{f}F_{\mu\nu}^a\tilde F_a^{\mu\nu}$ where $\frac{\lambda}{f}\sim\frac{\mathcal{O}(10)}{\mpl}$. It has a negligible effect on the background evolution, $\frac{\rho_{_{\rm YM}}}{\mpl^2H^2}\lesssim\epsilon^2$. However, its quantum fluctuations make a significant contribution to the cosmic perturbation. 
In particular, the gauge field has a spin-2 fluctuation which explicitly breaks the parity between the left- and right-handed polarization states. The chiral tensor modes are linearly coupled to the gravitational waves and lead to a circularly polarized tensor power spectrum comparable to the unpolarized vacuum power spectrum.
Moreover, the scalar sector is modified by the linear scalar fluctuations of the gauge field. Since the spin-0 and spin-2 fluctuations of the SU(2) gauge field are independent, the gauge field can, at the same time, generate a detectable chiral gravitational wave signal and have a negligible contribution to the scalar fluctuations, in agreement with the current CMB observations.}

\usepackage{xcolor,colortbl}
\definecolor{darkred}{rgb}{0.7,0,0}
\newcommand{\be}{\begin{equation}}
\newcommand{\bse}{\begin{subequations}}
\newcommand{\ese}{\end{subequations}}
\newcommand{\bea}{\begin{eqnarray}}
\newcommand{\eea}{\end{eqnarray}}
\newcommand{\ba}{\begin{array}}
\newcommand{\ea}{\end{array}}
\newcommand{\ee}{\end{equation}}
\newcommand{\nn}{\nonumber}

\def\dre_g{\delta\rho_g}
\def\dpe_g{\delta P_g}
\def\dqe_g{\delta q_g}
\def\dre{\delta\rho}
\def\dpe{\delta P}
\def\dqe{\delta q}
\def\tM{\tilde M}
\def\mH{\mathcal{H}}

\def\YM1{\frac{\dot\phi^2}{a^2}}
\def\YM2{\frac{g^2\phi^4}{a^4}}

\newcommand{\mpl}{M_{\rm pl}}

\def\dc{\delta\varphi}
\def\dd{\delta\psi}
\def\x{\tilde{\tau}}
\def\th{\tilde{h}}
\def\tg{\tilde{\gamma}}
\DeclarePairedDelimiter{\evdel}{\langle}{\rangle}
\newcommand{\ev}{\evdel}

\begin{document}

IPM/P-2016/003

\maketitle

\flushbottom

%\tableofcontents

\section{Introduction}

Cosmic inflation is a successful, well-studied paradigm which offers an elegant solution to many cosmological problems \cite{Guth:1980zm}. Besides, cosmological perturbations resulting from quantum fluctuations during inflation generate the seeds of the structures which we observe today. While many key predictions of inflation have been verified by CMB and LSS observations, still the primordial gravitational waves or B-mode polarization remains elusive \cite{Ade:2015lrj}. In 2014, the lensing B-mode signal has been directly detected by Polarbear \cite{Ade:2014afa} and shortly after, BICEP2 \cite{Ade:2014xna} pushed its constraints to a level that is competitive with temperature. The current upper limit on tensor fluctuations ($r_{0.05} < 0.07$ at $95 \%$ CL) comes from the latest joint analysis of Planck and BICEP2/Keck array measurements \cite{Array:2015xqh}. We are living in the golden age of observational cosmology and the quest for inflationary gravitational waves is the major goal of several observational projects. The road ahead seems promising for the detection of primordial gravitational waves and the discovery of new physics underlying inflation \cite{Kovetz:2015pia, Creminelli:2015oda, Schwarz:2015cma}. In case of single scalar field scenarios of inflation, by observing the primordial gravitational wave, we can determine both the energy scale of inflation, $V^{\frac14}\simeq10^{16}\textmd{Gev}\big(\frac{r}{0.01}\big)^{\frac14}$, and the inflaton field excursion, $\Delta\varphi\gtrsim\big(\frac{r}{0.01}\big)^{\frac12}\mpl$ \cite{Lyth:1996im}. However, that relations can in principle be evaded in cases that the gravitational waves are coupled to some new fields during inflation which has a negligible contribution to the scalar sector.

Axion fields are abundant in string theory and therefore very well-motivated candidates for the inflaton field. Enjoying shift symmetry, their effective potential is protected from dangerous quantum corrections which guaranteed the flatness of the potential. 
The axion field, $\varphi$, is classically coupled to gauge fields through a topological term $F\tilde F$, which is hence invariant under shift transformations of the form $\varphi\to \varphi +\varphi_0$ for an arbitrary $\varphi_0$ shift. On the other hand, quantum effects (\textit{i.e.} instanton contributions) induce a perturbatively exact cosine-type potential for the axion $V(\varphi)= \mu^4 (1+\cos(\varphi/f))$ which breaks the continuous shift symmetry to the discrete symmetry of $\varphi\to \varphi+2\pi f$ \cite{Weinberg-QFT-II}. Here, $\mu$ is the scale of the (approximate) shift symmetry breaking and $f$ is the axion decay constant. Since super-Planckian axion decay constant is hard to realize in string theory \cite{Svrcek:2006yi, Banks:2003sx}, the axion potential is under theoretical control if $H\!<\!f\!<\!M_{\rm pl}$. The lower limit on $f$ comes from the fact that the axion theory arises from integrating out modes heavier than $f$, hence, it can only work in inflation scales lower than that. For an exhaustive review of axion inflation see \cite{Pajer:2013fsa} and a comprehensive survey of axion inflation in string theory is presented in \cite{Long:2016jvd}.

The first model of axion inflation has been proposed more than 25 years ago in \cite{Freese:1990rb} and called \textit{natural inflation}.  Although natural inflation could rectify the naturalness problem by means of the shift symmetry and radiative stability of the potential,  does not fully resolve it. In fact, to have a successful inflationary background, this model needs a super-Planckian $f$ parameter which is not a natural scale within particle physics models. Natural inflation is now disfavoured by the joint BICEP2/Keck Array and Planck data. One of the most popular and well-motivated axion models of inflation is \textit{monodromy inflation} \cite{Silverstein:2008sg,Flauger:2009ab, McAllister:2014mpa, Easther:2013kla, Flauger:2014ana}. This inflationary mechanism is a string theoretic construction based on a single axion field and motivates a broad class of axion potentials of the form $V(\varphi)=\mu^{4-p}\varphi^p+\Lambda^4 e^{-c(\frac{\varphi}{\varphi_0})^{p_{\Lambda}}}\cos\bigg(\frac{\varphi_0}{f}(\frac{\varphi}{\varphi_0})^q+\theta_0\bigg)$. While the underlying periodicity of the theory continues to protect the inflaton potential from corrections, the periodic field space of the axion is now effectively unfolded due to the monodromy.

Besides their appealing theoretical stability, models of axion inflation are attractive phenomenologically due to their ability to generate observable primordial gravitational waves. These models can create detectable gravitational waves either as vacuum fluctuations of a large field model or sourced perturbations through their interaction with the gauge fields. Axions can naturally couple to gauge fields, Abelian or non-Abelian, and creates a richer phenomenology which leads to new observational and theoretical features. One possible construction is an axion driven inflation which interacts with a U(1) gauge field via $\varphi F\tilde F$. The \textit{Abelian} gauge field quanta is mixed to the gravitational waves at the \textit{nonlinear level} through the interaction $\delta A+\delta A\rightarrow\delta g$. That mechanism generates sourced chiral gravitational waves in addition to the standard (unpolarized) vacuum fluctuations \cite{Sorbo:2011rz}. However, the U(1) gauge field quanta is also coupled to the inflaton via $\delta A+\delta A\rightarrow\delta\varphi$ and generates large amounts of non-Gaussianity. In other words, the resulting sourced gravity wave signal is correlated to the large scale non-Gaussianity. Therefore, once the CMB constraints are imposed, the gravitational waves sourced by the U(1) gauge field are undetectable \cite{Barnaby:2010vf, Barnaby:2011qe, Barnaby:2011vw}. Authors of \cite{Namba:2015gja} evades that issue by considering an inflationary scenario in which the U(1) gauge field is coupled to a fast rolling axion field while both fields are only gravitationally coupled to the inflaton field.

Another natural possibility to study as the matter content of axion inflation is a (dark) SU(2) gauge field, $A^a_{~\mu}$. Thanks to the SU(2) algebra in such scenarios, there exists a homogeneous and isotropic field configuration for the gauge field \cite{Maleknejad:2011jw, Maleknejad:2011sq, Maleknejad:2012fw}. Therefore, the mixing between the \textit{non-Abelian} gauge field and perturbations in the scalar and tensor sectors are at the \textit{linear order} and coming from different fluctuations. Hence, the enhancement of gravitational wave and the modification in the scalar perturbations are uncorrelated.  One of the possible realizations of axion inflationary models involving non-Abelian gauge fields is \textit{chromo-natural inflation} \cite{chromo-natural-short}. 
In this model, the axion has a standard cosine potential and is coupled to the gauge field with $-\frac{\lambda}{4f}\textmd{tr}(F^a_{\mu\nu}F_a^{\mu\nu})$. The gauge field has an energy density $\rho_{_{\rm YM}}\sim \epsilon \mpl^2H^2$ and $\frac{\lambda}{f}\sim\frac{\mathcal{O}(10^3)}{\mpl}$ which leads to slow-roll inflationary background, without requiring super-Planckian $f$ \cite{Adshead:2012qe, Martinec:2012bv, Maleknejad:2012dt, Adshead:2013qp}. 
%In particular converting most of the potential energy of axion into the gauge field energy, the Chern-Simons interaction slows down the speed of the axion field rolling down its potential 
Moreover, the tensor fluctuations of gauge field source a chiral spectrum of gravitational waves. Despite its technical naturality, chromo-natural inflation has been disfavored by Planck data \cite{Dimastrogiovanni:2012ew, Adshead:2013nka}. In particular, the scalar perturbations of the model are stable if the magnetic to electric ratio of the vev gauge field is more than $\sqrt{2}$, and it is otherwise unstable. The source of instability in the scalar sector is coming from the interaction term $\frac{\lambda}{f}\big(\frac{\rho_{_{\rm YM}}}{H^2}\big)^{\frac12}\frac{1}{k\tau}$ which gets relevant at the intermediate regime $-k\tau=\frac{\lambda}{f}\big(\frac{\rho_{_{\rm YM}}}{H^2}\big)^{\frac12}\sim\mathcal{O}(10^{2})$.
 The tensor perturbations are however enhanced at large magnetic to electric ratio. Therefore, depending on the parameters, this model can either overgenerate gravitational waves or predicts a too red spectral tilt \cite{Adshead:2013nka, Obata:2016tmo}.

In this paper, we focus on a single field axion inflation in the presence of an SU(2) gauge field with a small vev ($\rho_{_{\rm YM}}\lesssim\epsilon^2\mpl^2H^2$). For the sake of generality, here we consider an arbitrary potential for the axion that is able to support the slow-roll inflation. The gauge field is coupled to the axion through a Chern-Simons interaction $-\frac{\lambda}{4f}\textmd{tr}(F^a_{\mu\nu}F_a^{\mu\nu})$ with $\frac{\lambda}{f}\sim\mathcal{O}(10)$. This interaction with the gauge field is expected as it is compatible with all the symmetries of the axion. Moreover, due to the SU(2) algebra, the gauge field can have an isotropic and homogeneous field configuration. It has a negligible effect on the background evolution as $\rho_{_{\rm YM}}\lesssim\epsilon^2\mpl^2H^2$ and the coupling between the gauge field and the axion is small. The quantum fluctuations of the gauge field, however, makes a significant contribution to the cosmic perturbation. In particular, the spin-2 fluctuations of the perturbed gauge field linearly coupled to the primordial gravitational waves and explicitly breaks the parity between the left- and right-handed polarization states. Therefore, our gravity waves has a circularly polarized power spectrum proportional to $\frac{\rho_{_{\rm YM}}}{\mpl^2H^2}$ which can be comparable to the power spectrum of its vacuum fluctuations. That results in parity odd CMB correlations between E and B-modes and T and B-models. Moreover, the perturbed gauge field has some scalar degrees of freedom which are linearly coupled to the curvature perturbations via $\frac{\lambda}{f}\big(\frac{\rho_{_{\rm YM}}}{\mpl^2H^2}\big)^{\frac12}\frac{1}{k\tau}$. In this scenario, the interaction terms are more relevant after horizon crossing, $-k\tau\sim\mathcal{O}(0.1)$. Therefore, the scalar sector is modified by the SU(2) gauge field at large scales. Our scalar perturbations are stable and almost adiabatic in case that the background magnetic to electric ratio of the gauge field is more than $\sqrt{2}$ while otherwise deviates from the adiabatic solution. There are parameter regimes in which the gauge field, at the same time, generates a detectable chiral gravitational wave signal and has a negligible contribution to the scalar fluctuations, in agreement with the current CMB observations. Hence, it satisfies in a modified version of the Lyth bound and the tensor power spectrum does not specify the scale of inflation.

%-----------------------------------------

This paper is organized as follows. Section \ref{basic-setup} presents the basic setup of the model. In section \ref{perturbation}, we classify its cosmic perturbation theory and work out the field equations. The scalar and tensor perturbations are studied in section \ref{Scalar perturbations} and \ref{tensor-section} respectively. Finally, we summarize in section \ref{conclusion}. Some technical details are presented in appendices A and B.

\section{Theoretical setup}\label{basic-setup}

We consider a generic axion-driven inflation model with a gauge field sector, both minimally coupled to Einstein gravity 
\bea\label{action}
\mathcal{L}_{\textmd{inf}} =\frac{R}{2}-\frac12\partial_\mu\varphi\partial^\mu\varphi-V(\varphi)+\mathcal{L}_{A}(A^a_{\mu},g_{\mu\nu},\varphi)\,,
\eea
where $\varphi$ is the axion field, $V(\varphi)$ is the axion potential and $\mathcal{L}_{A}$ is the gauge field sector. Here and throughout, the reduced Planck mass is set to unity, unless otherwise specified. For the purpose of this work and in order to be as model-independent as possible, $V(\varphi)$ is an arbitrary potential that is able to support the slow-roll inflation. In addition to the inflaton, we have a SU(2) gauge field which through the Chern-Simons interaction couples to the axion field
\bea\label{action-A}
\mathcal{L}_{A}(A^a_{\mu},g_{\mu\nu},\varphi)=-\frac{1}{4}\bigg(F^a_{\mu\nu}F_a^{\mu\nu}+\frac{\lambda}{f}\varphi\ F^a_{\mu\nu}\tilde{F}_a^{\mu\nu}\bigg)\,,
\eea
where $\lambda$ is a dimensionless parameter, $f$ is the axion decay constant and $\tilde{F}^{a\mu\nu}=\frac12\epsilon^{\mu\nu\lambda\sigma}F^a_{\lambda\sigma}$.
 The gauge field strength tensor is
\be
F^a_{~\mu\nu}=\partial_\mu A^a_{~\nu}-\partial_\nu A^a_{~\mu}-g\epsilon^a_{~bc}A^a_{~\mu}A^b_{~\nu},
\ee
where $g$ is the gauge coupling, $a, b, c...$ are the indices of the $su(2)$ algebra with generators $\{T_a\}$, defined by the commutation relation  $[T_a,T_b]=i\epsilon_{ab~}^{~~c}T_c$.

\subsection{Geometry of the isotropic configuration}

In the flat FLRW metric
\be
\label{FLRW}
ds^2=-dt^2+a(t)^2\delta_{ij}dx^{i}dx^{j},
\ee
and after choosing the temporal gauge for the gauge field $(A^a_0=0)$, we have the following isotropic and homogeneous field configuration 
\be\label{ansatz}
\varphi=\varphi(t)  \quad \textmd{and}\quad A^a_{~\mu}(t)= \psi(t)e^a_{~\mu},
\ee%
%\be\label{ansatz}
%\varphi=\varphi(t)  \quad \textmd{and}\quad A^a_{~\mu}(t)=\left\{
%\begin{array}{ll} \psi(t)a\delta^a_i\, ,\qquad  &\mu=i
%\\   0\,, \qquad &\mu=0\,
%\end{array}\right.
%\ee%
where $\{e^{\alpha}_{~\mu}\}$ are tetrads of FRW metric (with $e^{a}_{~0}=0$) and the effective field value of the gauge field $\psi$ is a pseudo-scalar. The tetrad fields are the noncoordinate orthonormal basis satisfying
\be
g_{\mu\nu}=e^{\alpha}_{~\mu}e^{\beta}_{~\nu}\eta_{\alpha\beta},
\ee
where $\alpha,\beta=0,1,2,3$ and $\eta_{\alpha\beta}$ is the Minkowski metric. For the FRW metric, $\{e^{\alpha}_{~\mu}\}$ are specified as 
\be
e^{0}_{~\mu}=n_{\mu} \quad \textmd{and} \quad e^{a}_{~\mu}=a(t)\delta^a_{\mu} \quad a=1,2,3,
\ee
where $n^{\mu}=(1,0,0,0)$ is the 4-velocity of the comoving observer.

The reason for the existence of such a homogeneous and isotropic solution is as follows \cite{Maleknejad:2011jw, Maleknejad:2011sq}.
Working in the temporal gauge $A^a_0=0$, under the action of an infinitesimal rotation $R(\vec{\theta})=e^{\vec\theta.\vec{M}}$, $A^a_i$ transforms as
\be\label{trans-R}
A^a_i\xmapsto{R} (R(\vec{\theta})A^a)_i=(\delta^j_i-\theta_k\epsilon_i^{~jk})A^a_{j},
\ee
where $M_i$s are generators of $SO(3)$ in 3-dimensional vector space, $(M_i)_{jk}=-\epsilon_{ijk}$.
On the other hand, setting $A^a_0=0$, only fixes $A^a_i$ up to global SU(2) gauge transformations\footnote{Under the action of a generic (local) gauge transformation $\Lambda(\lambda(t,\textbf{x}))=e^{i\lambda_aT^a}$, the gauge field transforms as $A_{\mu}\mapsto A_{\mu}-\frac{i}{g}\Lambda^{-1}D_{\mu}\Lambda$, where $D_{\mu}=\partial_{\mu}+ig A_{\mu}$ is the covariant derivative.} of the form $\Lambda(\lambda)=e^{i\lambda_aT^a}$. The residual (global) gauge transformation is in the form
\be\label{trans-Gauge}
A^a_i\xmapsto{\Lambda} \big(\Lambda^{-1}(\vec{\lambda})A_i\Lambda(\vec{\lambda})\big)^{\!a}=(\delta^a_b-\lambda^c\epsilon^{a}_{~bc})A^b_{i}=R(\vec{\lambda})^a_{~b}A^b_{i}.
\ee
From the combination \eqref{trans-R} and \eqref{trans-Gauge} we find that for all $\theta_k$s there exists a $\lambda_c=-\delta^k_c\theta_k$, so that $A^a_i\propto e^a_{~i}$ is invariant under the action of their combination. That then explains the existence of the isotropic and homogeneous configurations of the form \eqref{ansatz}. The isomorphism of $su(2)$ and $so(3)$ Lie algebras plays a key rule here and makes the identification of algebra and spatial indices of the local frame possible.

\subsection{Background evolution and slow-roll inflation}

The isotropic and homogeneous solution in \eqref{ansatz} gives the electric and magnetic field components as
\be
E^a_i=-(H\psi+\dot{\psi})\delta^a_i \quad \textmd{and} \quad B^a_i=-g\psi^2\delta^a_i.
\ee
The background energy densities of the axion and the gauge field are respectively
\bse
\begin{align}
\rho_{\varphi}&=\frac12\dot{\varphi}^2+V(\varphi),\\
\rho_{_{\rm YM}}&=\frac12\bigg(\vec{E}^a.\vec{E}_a+\vec{B}^a.\vec{B}_a\bigg).
\end{align}
\ese
The field equations of $\varphi$ and $\psi$ are
\bse\label{c-n.e.o.m}
\begin{align}
&\ddot\varphi+3H\dot\varphi+V_{\varphi}=-3\frac{\lambda g}{f} \psi^2(\dot\psi+H\psi)\,,\\\label{eq-psi}
&\ddot\psi+3H\dot\psi+(2H^2+\dot H)\psi+2g^2\psi^3=\frac{\lambda g}{f}\psi^2\dot\varphi\,,
\end{align}
\ese
which are coupled by the Chern-Simons interaction term.
Moreover, the continuity equations are
\bse
\begin{align}
&\dot\rho_{\varphi}+3H(\rho_{\varphi}+P_{\varphi})=-\frac{\lambda }{f}\dot{\varphi}\vec{E}^a.\vec{B}_a,\\\label{rho-ym}
&\dot\rho_{_{\rm YM}}+4H\rho_{_{\rm YM}}=\frac{\lambda }{f}\dot{\varphi}\vec{E}^a.\vec{B}_a.
\end{align}
\ese 
As we see explicitly in \eqref{rho-ym}, in the absence of the interaction term with the axion, $\rho_{_{\rm YM}}$ damps like $a^{-4}$. However, the Chern-Simons interaction breaks the conformal symmetry and prevents the damping of the gauge field (when $\dot\varphi\neq0$).

Considering the standard slow-roll inflation, we can quantify the slow-roll dynamics by 
\be\label{SL-H}
\epsilon\equiv-\frac{\dot H}{H^2}\quad  \textmd{and} \quad \eta\equiv-\frac{\ddot H}{2H\dot H}=-\frac{(\epsilon H^2\dot{)}}{2\epsilon H^3}.
\ee
We also demand the gauge field to have a slow varying evolution, therefore from \eqref{eq-psi} we realize that the dimensionless time derivatives of $\psi$
\be\label{vartheta}
\epsilon_{\psi}\equiv \frac{\dot\psi}{H\psi} \quad \textmd{and}  \quad \eta_{\psi}\equiv-\frac{\ddot{\psi}}{H\dot{\psi}},
\ee
should also be very small during slow-roll inflation.
It is useful to define two new parameters 
\be
\xi\equiv\frac{\lambda\dot{\varphi}}{2fH} \quad \textmd{and} \quad \xi_{\psi}\equiv\frac{B}{E},
\ee
where $E=(\vec{E}^a.\vec{E}_a)^{\frac{1}{2}}$ and $B=(\vec{B}^a.\vec{B}_a)^{\frac{1}{2}}$.
The ratio of the energy of dark radiation to total energy is
\be
\frac{\rho_{_{\rm YM}}}{\rho}\simeq\frac{\psi^2}{2}(1+\xi_{\psi}^2),
\ee
in which we neglect the sub-dominant term $\epsilon_{\psi}$. Hereafter, a $``\simeq$'' means up to the dominant order in slow-roll.

In our model, we are interested in the regime that
\be
\frac{\rho_{_{\rm YM}}}{\rho}\lesssim\epsilon^2,
\ee
thus, our slow-roll parameters are 
\be\label{SL-H-SL}
\epsilon\simeq\frac12\frac{\dot\varphi^2}{H^2}\quad \textmd{and} \quad \eta\simeq-\frac{\ddot\varphi}{H\dot\varphi}.
\ee
Up to the dominate order in slow-roll, we have  $\xi_{\psi}\simeq\frac{g\psi}{H}$ and $\xi\simeq\sqrt{\frac{\epsilon}{2}}\frac{\lambda}{f}$ which are  related as
\be\label{xi-eq}
\xi\simeq\frac{(1+\xi_{\psi}^2)}{\xi_{\psi}}.
\ee
During the slow-roll inflation, the energy density of the gauge field is almost constant and $\rho_{_{\rm YM}}\simeq\frac{\xi}{2}\vec{E}^a.\vec{B}_a$. For a $\xi\sim1$, we have $\xi_{\psi}\sim1$, $\frac{\lambda}{f}\sim1/\sqrt{\epsilon}$ and $\psi\sim\epsilon$. Since the large coupling is hard to achieve in a controlled string compactification \cite{Baumann:2014nda}, we are interested in small $\lambda$, e.g. $f\sim0.01$ and $\lambda\sim0.1$. As the axion rolls down its potential, $\dot{\varphi}/H$ increases and part of the energy of the axion gradually injects to the gauge field, therefore $\rho_{_{\rm YM}}$ (as well as $\psi$ and $\xi_{\psi}$) slowly increases during inflation. After the end of inflation on the other hand, $\dot{\varphi}$ starts oscillating around the minimum of the potential and the gauge field acts like a dark radiation sector, \textit{i.e.} $A^a_{i}\propto a^{-1}$.

\section{Cosmic perturbation theory}\label{perturbation}

In this section, we work out the cosmic perturbation theory of the axion model \eqref{action} in the presence of an SU(2) gauge field.
We are interested in linear perturbations in this paper. At the perturbation level, fields are perturbed around the isotropic and homogeneous configuration \eqref{ansatz}. Due to the quantum fluctuations, all the non-Abelian gauge field modes are turned on and can contribute to the perturbation theory. Dealing with non-Abelian gauge fields bring new features and complications compared to the standard axion
scalar models. However, because of the isotropy of the background, one can still use the scalar, vector, and tensor decomposition for the perturbations  \cite{Maleknejad:2012fw}.

\subsection{Classification of the fluctuations}
In this subsection, we turn to classify the field and metric fluctuations around the homogeneous and isotropic background solution.
The most general form of the perturbed FRW metric can be parametrized as
\be\label{metric-pert}%
ds^2=-(1+2A)dt^2+2a(\partial_iB+V_i)dx^idt+a^2\left((1-2C)\delta_{ij}+2\partial_{ij}E+2\partial_{(i}W_{j)}+\gamma_{ij}\right)dx^idx^j\,,
\ee
where $\partial_i$ denotes partial derivative respect to $x^i$ and $A,\ B,\ C$ and $E$ are scalar perturbations, $V_i,\ W_i$ parametrize vector perturbations (these are divergence-free three-vectors) and $\gamma_{ij}$, which is symmetric, traceless and divergence-free, is the tensor mode. The axion and the SU(2) gauge field are also perturbed around their homogeneous and isotropic background configurations (Eqn. \eqref{ansatz})
\be\label{field-perturb}
\varphi(t,\textbf{x})=\varphi(t)+\delta\tilde{\varphi}(t,\textbf{x})\quad \textmd{and} \quad A^a_{~\mu}(t,\textbf{x})=\left\{
\begin{array}{ll} a\psi(t)\delta^a_i+\delta A^a_{~i}(t,\textbf{x})\, ,\qquad  &\mu=i
\\   \delta A^a_{~0}(t,\textbf{x})\,, \qquad &\mu=0\,
\end{array}\right.
\ee%
where (as explained in appendix \ref{gauge-invariant}) the 12 components of $\delta A^a_{~\mu}(t,\textbf{x})$  are%
\bse \label{gauge-field-pert}
\begin{align}
\delta A^a_{~i}&=a\delta^a_i (\delta\psi-\psi C)+\delta^{aj}\big(\partial_{ij}(\tilde{Z}+a\psi E)+\partial_i (v_j+a\psi W_j)+a(\tg_{ij}+\frac{\psi}{2}\gamma_{ij})\big)\\
&+\epsilon^{a~j}_{~i}\big(ga\psi\partial_{j}(Z-\tilde{Z})+w_j\big),\nonumber\\
\delta A^a_{~0}&=\delta^{k}_a\partial_k(Y+a\psi\dot{E})+\delta_a^j (u_j+\psi V_j).
\end{align}
\ese
Because of the gauge transformations generated by space-time diffeomorphisms as well as the gauge transformations of $A^a_{\mu}$, not all the above 23 metric and fields perturbations are physically meaningful. Eliminating all the gauge symmetries, 4 coordinate freedoms and 3 internal gauge transformations, we then can construct 16 gauge invariant degrees of freedom. 
\begin{itemize}
\item{
On the \emph{scalar} sector, one can construct \emph{six} independent gauge-invariant combinations, two standard Bardeen potentials, the perturbed axion field and three gauge invariant combinations coming from the gauge field fluctuations
\be\label{scalar-gin}
\begin{array}{ll} \Psi=C+a^2H(\dot{E}-\frac{B}{a})\\
\Phi=A-\frac{d}{dt}\left(a^2(\dot{E}-\frac{B}{a})\right)\\
\end{array}
\quad \textmd{and}\quad
\begin{array}{llll} \dc=\delta\tilde\varphi-\dot\varphi a^2(\dot{E}-\frac{B}{a}),\\
\dd=\dd,\\
 M=g^2\psi^3aZ,\\
\tilde{M}=H\psi(\dot{\tilde{Z}}-Y).\,
\end{array}
\ee%
}
\item{There are \emph{three} gauge invariant divergence-free \emph{vector} perturbations, one from the metric fluctuation and two from the gauge field perturbations
\be\label{vector-gin}
\mathcal{Z}_i=a\dot{W}_i-V_i\,, \quad \textmd{and} \quad \begin{array}{ll} \mathcal{U}_i=\frac1g\dot{w}_i+u_i, \\
 \mathcal{V}_i=\frac1g w_i+v_i.
 \end{array}
\ee
}
\item{On the \emph{tensor} sector, we have two tensor perturbations $\gamma_{ij}$ and $\tg_{ij}$, which are both gauge invariant with two degrees of freedom.  The tensor perturbations are, by definition, symmetric, traceless and divergence-free.}
\end{itemize}

\subsection{Independent field equations}
Working out the gauge-invariant combinations, we are now ready to field the linearized field equations that govern their
dynamics.
The linear order perturbed energy-momentum tensor around a background perfect fluid can be decomposed as %
\begin{align}
\delta T_{ij}=&\bar P\delta g_{ij}+a^2\left(\delta_{ij}(\delta P-\frac13\nabla^2\pi^S)+\partial_{ij}\pi^S
+\partial_i\pi^V_j+\partial_j\pi^V_i+\pi^T_{ij}\right)\,,\cr
\delta T_{i0}=&\bar P\delta g_{i0}-(\bar \rho+\bar P)(\partial_i\delta u+\delta u_i^V)\,,\cr
\delta T_{00}=&-\bar\rho\delta g_{00}+\delta \rho\,,\nn
\end{align}
where $\bar\rho$ and $\bar P$ are the background energy and pressure densities. Moreover,  $\pi^S$, $\pi^V_i$, $\pi^T_{ij}$ represent the \textit{anisotropic inertia},
characterizing departures from the perfect fluid form of the
energy-momentum tensor, while $\delta u_i^V$ is the vorticity. They satisfy the following conditions
$$\partial_i\pi^V_i=\partial_i\pi^T_{ij}=\partial_i\delta u_i^V=0.$$

One can construct the following four gauge invariant combinations from $\delta\rho$, $\delta P$ and $\delta q$
\bse\label{diff-inv-Tmunu}
\begin{align}
\delta\rho_g=&\delta\rho-\dot{\bar \rho}a^2(\dot{E}-\frac{B}{a})\,,\nonumber\\
\delta P_g=&\delta P-\dot{\bar P}a^2(\dot{E}-\frac{B}{a})\,,\nonumber\\
\delta q_g=&\delta q+(\bar\rho+\bar P)a^2(\dot{E}-\frac{B}{a})\,,\nonumber
\end{align}\ese
while  $\pi^S$, $\pi^V_i$, $\pi^T_{ij}$ and $\delta u_i^V$ are gauge invariant quantities, where $\delta q=(\bar\rho+\bar P)\delta u$.
It is useful to decompose the energy-momentum tensor into the contribution of the axion and the gauge field as $$\delta T^{\mu\nu}=\delta T^{\mu\nu}_{\varphi}+\delta T^{\mu\nu}_{_{\rm YM}}.$$ The axion sector, $\delta T^{\mu\nu}_{\varphi}$, is specified by
\bse
\begin{align}
%\label{delqe}
\dqe_{\varphi}&=-\dot{\varphi}\dc,\\
\dre_{\varphi}&=\dot\varphi\delta\dot\varphi-\dot\varphi^2\Phi+V_{\varphi}\dc,\\
%\label{delta-P}
\dpe_{\varphi}&=\dot\varphi\delta\dot\varphi-\dot\varphi^2\Phi-V_{\varphi}\dc,
\end{align}
\ese
while $\delta T^{\mu\nu}_{_{\rm YM}}$ has the following momentum, energy and pressure densities 
\bse
\begin{align}
%\label{delqe}
\dqe_{_{\rm YM}}&=-2\dot M+2H\bigg(M+\xi^2_{\psi}\tM-\psi\dd+\psi^2\Psi\bigg),\\
\dre_{_{\rm YM}}&=3H^2\psi^2\bigg(\frac1H(\frac{\delta\psi}{\psi}\dot{)}-\Phi+(1+2\xi_{\psi}^2)\frac{\dd}{\psi}\bigg)-\frac{k^2}{a^2}(\tM+2M),\\
%\label{delta-P}
\dpe_{_{\rm YM}}&=\frac13\dre_{_{\rm YM}}.
\end{align}
\ese
Unlike the axion energy-momentum tensor, $\delta T^{\mu\nu}_{_{\rm YM}}$ deviates from the perfect fluid form. In other words, although the background energy-momentum tensor is in the form of a perfect fluid, at the perturbation level, $\delta T^{\mu\nu}_{_{\rm YM}}$ is an \textit{imperfect fluid} with non-vanishing anisotropic inertia and vorticity as
\bse\label{devi}
\begin{align}
\label{piS}
a^2\pi^S&=2(M-\tM),\\
a\label{piv}\pi_i^V&=H\psi\bigg(H\xi^2_{\psi}\mathcal{V}_i+(\mathcal{U}_i-\dot{\mathcal{V}}_i-\psi\mathcal{Z}_i)\bigg)\,,\\
\label{piT}
\pi^T_{~ij}&=2H\psi\bigg((\xi^2_{\psi}-1)H\tg_{ij}-\dot{\tg}_{ij}-\frac{\psi}{2}\dot{\gamma}_{ij}
+\xi_{\psi}\partial_k\epsilon^{kl}_{~~(i}\big[\tg_{j)l}+\frac{\psi}{2}\gamma_{j)l}\big]\bigg)\,,\\
\delta q^V_i&=
H\psi\bigg(\xi_{\psi}\nabla\times\big(\dot{\vec{\mathcal{V}}}-\vec{\mathcal{U}}+\psi\vec{\mathcal{Z}})-2\xi^2_{\psi}H\vec{\mathcal{U}}
-\xi_{\psi}H(\nabla\times\vec{\mathcal{V}})\bigg)_i.
\end{align}
\ese

As follows from \eqref{pi^s}-\eqref{dP}, \eqref{firstV}-\eqref{delta-q-V} and \eqref{T-gf}, there are ten independent Einstein equations, \emph{four} scalars, \emph{two} vectors and \emph{one} tensor. Since they are less than the number of (physical) gauge-invariant quantities, one needs more equations to have a complete set of equations. These extra equations are provided by the field equations which are given by the second order action. In fact, the scalar and vector parts of the gauge field equations
can be written as\footnote{These extra equations are the field equation of $A^a_{~0}$ component which are constraints
enforcing the gauge invariance of the action. Note that dealing with a gauge invariant action, $\dot A^a_{~0}$
does not appear in the Lagrangian density, $\mathcal{L}$, and the momentum conjugate to $A^a_{~0}$ is identically zero.}
\bse
\label{constraint-2nd-order-action}
\begin{align}
\delta^{k}_a\partial_k\big(\frac{\partial\delta\!_{_{2}}(\sqrt{-g}\mathcal{L})}{\partial Y }\big)&=0,\\
\delta^a_i\big(\frac{\partial\delta\!_{_{2}}(\sqrt{-g}\mathcal{L})}{\partial u_i }\big)&=0\,,
\end{align}
\ese
where $\delta_2$ stands for second order in perturbations. The equation of motion for the perturbed axion field $\dc$ and the tensor mode $\tg_{ij}$ will also be
obtained from the corresponding parts of the second order action. In the following table, we summarize the number of gauge-invariant perturbations and the independent equations governing the dynamics of each part of the system.
\begin{center}
\begin{tabular}{cp{1.5cm}p{3cm}p{3cm}p{3cm}p{2cm}}
%& & \multicolumn{5}{}{Fields} & \multicolumn{3}{}Equations \\
\hline
&  & \begin{footnotesize}
Gauge-invariants
\end{footnotesize} &\begin{footnotesize}
Einstein Eqn.s
\end{footnotesize}  & \begin{small}
($\frac{\delta S}{\delta A})_{_{_{\!(\!1\!)}}}$
\end{small}  & \begin{small}
$(\frac{\delta S}{\delta \varphi})_{_{_{\!(\!1\!)}}}$
\end{small}\\ [2.5ex]
\hline
 &\begin{footnotesize}
Scalar
\end{footnotesize}  & 6 & 4 & 1 & 1    \\ [1.1ex]
& \begin{footnotesize}
Vector
\end{footnotesize}  & 3 & 2 & 1 & 0    \\[1.1ex]
& \begin{footnotesize}
Tensor
\end{footnotesize}  & 2 & 1  & 1 & 0    \\ [1.1ex]  \cline{1-6}
\end{tabular}
\vskip 0.1 cm
\textbf{Table I: Gauge-invariant perturbation modes and independent field equations}
\end{center}
In the table I, $(\frac{\delta S}{\delta A})_{_{_{\!(\!1\!)}}}$ and $(\frac{\delta S}{\delta \varphi})_{_{_{\!(\!1\!)}}}$ represent the linear order field equations of the gauge field and the axion field which are determined by the second order action. Here, we only present the final results, for more details we refer to \cite{Maleknejad:2012fw}.

%------------------------------------------------------

For later convenience, here we introduce two Fourier space variables in terms of conformal time $\tau$ and comoving momentum $k$
\be\label{T-def}
\x\equiv-k\tau \quad \textmd{and} \quad \tilde\mH\equiv\frac{\mH}{k},
\ee
where $\mH=aH$. During the slow-roll inflation in which $\mH\simeq-(1+\epsilon)/\tau$, we have
\be
\x\simeq\frac{k_{\rm phy}}{H}\quad \textmd{and} \quad \tilde\mH\simeq\frac{(1+\epsilon)}{\x}.
\ee
in which $k_{\rm phy}$ is the physical momentum $k/a$.

\subsubsection{Scalar sector}

In the scalar sector of the perturbations, we have \emph{six} gauge-invariant combinations of \eqref{scalar-gin}, $\{\delta\varphi,\dd,M,\tM,\Psi,\Phi\}$. These perturbations are governed by \emph{four} scalar Einstein equations, the field equation of $\delta A^a_{~0}$ ( Eqn. \eqref{constraint-2nd-order-action}) and $\dc$.

The scalar part of the perturbed Einstein equations take the form
\bse%
\begin{align}\label{pi^s}
&a^2\partial_{ij}\pi^s=\partial_{ij}(\Psi-\Phi)\,,\\
\label{dq} &\partial_{i}(\dqe_g+2(\dot{\Psi}+H\Phi))=0\,,\\
\label{drho} &\dre_g-3H\dqe_g+2\frac{k^2}{a^2}\Psi=0\,,\\
\label{dP}
&\dpe_g+\dot{\dqe}_g+3H\dqe_g+2\epsilon H^2\Phi-
\frac23\frac{k^2}{a^2}(\Psi-\Phi)=0\,. %
\end{align}
\ese%
Moreover, the scalar part of the field equation of $\delta A^a_{~0}$ (Eqn. \eqref{constraint-2nd-order-action}) is the constraint below
\bea\label{A02}
H\dqe_g-H\psi^2\big(\frac{\dd}{\psi}\dot{\big)}+(\dot{\varphi}+\frac{\lambda g\psi^3}{f})H\dc+H^2\psi^2(\frac{\dd}{\psi}+\Phi)+\frac{k^2}{a^2}\tM=0.~~~~~~
\eea
The field equation of $\delta\varphi$ is
\bea\label{deltachi-eq}
\delta\ddot\varphi+3H\delta\dot{\varphi}+\bigg(\frac{k^2}{a^2}+V_{\varphi\varphi}\bigg)\dc=2(\ddot\varphi+3H\dot\varphi)\Phi+\dot\varphi(\dot\Phi+3\dot\Psi)-\frac{\lambda}{f}\delta(\vec{E}^a.\vec{B}_a)
\eea
where $\delta(\vec{E}^a.\vec{B}_a)$ is the linear order perturbation of $\vec{E}^a.\vec{B}_a$ which is
%\be
%\frac{\lambda}{8f}\delta(F\wedge F)=-\frac{3\lambda g\phi^2}{fa^3}\bigg(\dot Q+\frac{2\dot\phi}{\phi}Q+\dot\phi(3\Psi-\Phi)\bigg)+\frac{k^2}{a^2}\frac{\lambda}{f}\frac{a\dot\phi}{g\phi^2}\bigg(2M+\frac{g^2\phi^4}{a^2\dot\phi^2}\tM\bigg).
%\ee
\be\label{wedge}
\delta(\vec{E}^a.\vec{B}_a)=3 g\psi^3H\bigg(\frac{1}{H}\bigl(\frac{\dd}{\psi}\dot{\bigl)}+3\big(\frac{\dd}{\psi}\big)-\Phi-\frac{k^2}{3a^2}(\frac{2M}{g^2\psi^4}+\frac{\tM}{H^2\psi^2})\bigg).
\ee
Eqn.s \eqref{pi^s}-\eqref{dP}, \eqref{A02} and \eqref{deltachi-eq} provides enough number of equations for $\dc$, $\dd$, $\Psi$, $\Phi$, $M$ and $\tM$. In sec. \ref{Scalar perturbations}, we solve these equations and study scalar fluctuations during the slow-roll inflation.

\subsubsection{Vector sector}

The vector perturbations of the metric and the gauge fields have three gauge invariant combinations of Eqn.\eqref{vector-gin}, $\{\mathcal{V}_i,\mathcal{U}_i,\mathcal{Z}_i\}$. The perturbed Einstein equations involves two vector equations, one constraint and one dynamical equation, given as 
\bse 
\label{vector-eins}
\begin{align}
\label{firstV}
&\partial_i\left(2a^2\pi^V_j-\frac{1}{a}(a^2\mathcal{Z}_j\dot{)}\right)=0\,,\\\label{delta-q-V}
&2a\delta q_i^V+\nabla^2\mathcal{Z}_i=0\,.
\end{align}
\ese
Dealing with three unknowns, the last equation is provided by the vector part of the field equation of $\delta A^a_{~0}$.
Explicitly, using \eqref{delta-q-V} in the vector part of \eqref{constraint-2nd-order-action} yields to
%\be\label{2nd-vector}
%-2\frac{g^2\phi^3}{a^2}(\mathcal{U}_i+\frac{\phi}{a}\mathcal{Z}_i)
%+\frac{g\phi^2}{a^2}\big(\vec\nabla\times(\dot{\vec{\mathcal{V}}}
%+\frac{\phi}{a}\vec{\mathcal{Z}})\big)_i-\frac{g\phi\dot\phi}{a^2}(\vec\nabla\times\vec{\mathcal{V}})_i
%-\frac{\phi}{a^2}\nabla^2(\mathcal{U}_i-\dot{\mathcal{V}}_i)=0.
%\ee
\be\label{vec-const}
g\psi^2\vec\nabla\times\vec{\mathcal{U}}_i-
\frac{\psi}{a}\nabla^2(\mathcal{U}_i-\dot{\mathcal{V}}_i-\psi\mathcal{Z}_i)+\frac{1}{2a}\nabla^2{\cal Z}_i=0 \,.
\ee
This completes the set of equations we need for solving vector perturbations.
Then, the combination of \eqref{firstV}-\eqref{delta-q-V} and \eqref{vec-const} indicates that $\cal Z$ exponentially damps during inflation.
From the combination of  \eqref{devi} and \eqref{vector-eins}, we then find that $\mathcal{Z}_i$ vanishes
after horizon crossing. Despite having gauge fields in our matter content, the power spectrum of the vector modes are unimportant in inflationary cosmology and CMB anisotropies.

\subsubsection{Tensor sector}

In the tensor sector, we have two gauge invariant tensors each with two degrees of freedom: the spin-2 fluctuations of the metric $\gamma_{ij}$ (\textit{gravitational waves}) and the gauge field $\tg_{ij}$, which we call \textit{tensor waves}.
 These tensor modes are governed by the tensor part of the Einstein equation and the field equation of $\tg_{ij}$ given by the second order action. Tensor fluctuations of the SU(2) gauge field interact with the tensor perturbations of the metric and modify its linear order field equation. These new interactions in the quadratic action involve parity odd terms which generate chiral tensor modes. Here, we only focus on the tensor perturbations of the axion inflation in \eqref{action}. However, the above property is the generic feature of inflationary models in the presence of a non-Abelian gauge field \cite{Maleknejad:2014wsa}.

The perturbed Einstein equations involve one equation for $\gamma_{ij}$
\be \label{T-gf}
\ddot \gamma_{ij}+3H \dot \gamma_{ij}-\frac{\nabla^2}{a^2}\gamma_{ij}=2\pi^T_{ij}\,,
\ee
in which $\pi^T_{ij}$ is the tensor part of the anisotropic inertia\footnote{Comparing with the exact form of $\pi^T_{ij}$ in \eqref{devi}, here in \eqref{pi-T} we dropped two slow-roll suppressed derivatives of $\gamma_{ij}$ of the form $\frac{\rho_{\rm YM}}{H}\dot{\gamma}_{ij}$ and $\frac{\rho_{\rm YM}}{H}\epsilon^{kl}_{~~i}\partial_k\gamma_{jl}$. }
\be\label{pi-T}
\pi^T_{~ij}=2H\psi\bigg((\xi^2_{\psi}-1)H\tg_{ij}-\dot{\tg}_{ij}
+\xi_{\psi}\partial_k\epsilon^{kl}_{~~(i}\tg_{j)l}\bigg)\,,
\ee
Note that $\pi^T_{ij}$ is proportional to $\psi$, the effective field value of the gauge field in the background level. Therefore, in order to have a linear order anisotropic inertia, the gauge fields should be turned on at the background level.
Moreover, the field equation of the tensor perturbation of the gauge field $\tg_{ij}$
is provided by its second order action
\bea
\label{2ndts}
&&\hspace*{-7mm}\delta\!_{_{2}}S_{\tilde h}\simeq\frac12\int d^3x dt a^3\biggl(
\big(\dot{\tg}_{ij}\big)^2-\big(\frac{\partial_k\tg_{ij}}{a}\big)^2-2\xi\xi_{\psi}H^2\tg_{ij}^2+2(\xi
+\xi_{\psi})H\epsilon^{ijk}\tg_{kl}\frac{\partial_i\tg_{jl}}{a}\nn\\
&&~~~~~~+2H\psi\big(\dot{\gamma}_{ij}+\xi\epsilon^{ikl}\frac{\partial_k\gamma_{jk}}{a})\tg_{ij}\biggr)\,.
\eea
Interestingly, both $\gamma_{ij}$ and $\tg_{ij}$ have sound speeds equal to one. It is noteworthy to mention that the quadratic action above involves all the possible combinations of $\tilde\gamma\tilde\gamma$ with $n\leq2$ derivatives.
Among them, we have two parity violating terms, $\epsilon^{ijk}\tg_{kl}\partial_i\tg_{jl}$ and $\epsilon^{ijk}\tg_{kl}\partial_i \gamma_{jl}$, which are originated from the Yang-Mills and Chern-Simons terms in the action.

Going to the Fourier space, we can diagonalize the system in terms of circular polarizations. In terms of the right- and left-handed polarizations, $\gamma_{ij}$ and $\tg_{ij}$ are decomposed as
\bse
\begin{align}
&\gamma_{ij}(\tau,\textbf{x})=\frac{1}{\sqrt{2}a}\sum_{\sigma=R,L}\int \frac{d^3k}{(2\pi)^\frac32} h_{\sigma}(\tau,\textbf{k})e^{\sigma}_{ij}(\textbf{k})e^{i\textbf{k}.\textbf{x}},\\ &\tg_{ij}(\tau,\textbf{x})=\frac{1}{2\sqrt{2}a}\sum_{\sigma=R,L}\int \frac{d^3k}{(2\pi)^\frac32} \th_{\sigma}(\tau,\textbf{k})e^{\sigma}_{ij}(\textbf{k})e^{i\textbf{k}.\textbf{x}},
\end{align}
\ese
where $\{h_{_{R,L}},\th_{_{R,L}}\}$ are the canonically normalized fields and $e^{R,L}_{ij}$ are the circular polarization tensors which satisfy the conditions
\bse
\begin{align}
&e^{\sigma}_{ij}e^{\sigma'*}_{ij}=2\delta^{\sigma\sigma'},\\
&\epsilon^{ijk}\hat{k}_ie^{\sigma}_{kl}=i\lambda_{\sigma}e^{\sigma j}_{l}, \quad \textmd{with}  \quad \lambda_{_{R,L}}=\pm1.
\end{align}
\ese

For a wave vector $\textbf{k}=(0,0,k)$, the right- and left-handed modes are defined as $h_{_{R,L}}\equiv a(\gamma_{11}\pm i\gamma_{12})/2$.
From the second order action \eqref{2ndts}, we obtain the field equation of $\th_{_{R,L}}(\tau,\textbf{k})$ as
\bea\label{v-eq}
&&\th''_{_{R,L}}+\bigg(k^2\mp2(\xi+\xi_{\psi})k\mH+2\xi\xi_{\psi}\mH^2\bigg)\tilde{h}_{_{R,L}}\simeq2\psi\mH
\bigg(h'_{_{R,L}}-\mH h_{_{R,L}}\pm k\xi h_{_{R,L}}\bigg),\nonumber\\
\eea
in which we have parity odd terms that have different signs for the right- and left-handed polarizations. Using the slow-roll relation \eqref{xi-eq} in the above and recalling that $h_{_{R,L}}\propto a$, we realize that the RHS of \eqref{v-eq} vanishes in the long wavelength limit. 
In sec. \ref{tensor-section}, we solve the field equations of $\{h_{_{R,L}},\th_{_{R,L}}\}$ and study tensor fluctuations during the slow-roll inflation.

\section{Scalar perturbations}\label{Scalar perturbations}

In the scalar sector, we have six independent fields and six equations. 
Upon using variable redefinition \eqref{T-def}, it is straightforward to see that all of our equations can be written in terms of $\x$ and $\tilde{\mH}$. For instance, we can write the field equation of $\delta\varphi$ ( Eq. \eqref{deltachi-eq}) as
\bea\label{delta-varphi-}
&&(a\delta\varphi)_{\x\x}+\bigg(1-(2-3\eta-\epsilon)\tilde{\mathcal{H}}^2\bigg)a\delta\varphi\simeq\frac{6\dot{\varphi}}{H}\tilde{\mH}^2 a\Phi-\frac{3\lambda g\psi^3}{fH}\bigg(\tilde{\mH}^2(\frac{2a\delta\psi}{\psi}-a\Phi)-\tilde{\mH}\big(\frac{a\delta\psi}{\psi}\big)_{\x}\nonumber\\
&-&\frac{1}{3\psi^2}(\frac{2aM}{\xi_{\psi}^2}+a\tilde M)\bigg).
\eea
Assuming slow-roll inflation, all of the coefficients in our equations are slow varying with time and approximately constant up to the dominant order in slow-roll. Thus, all of our six fields are functions of $\x$ with a coefficient of $k$ which is given by the initial value. Setting the initial value of the canonically normalized fields by the standard Bunch-Davis, solutions has the following formal forms
$$X_{I}(\tau,k)=\frac{1}{\sqrt{k}}f_I(\x)\quad \textmd{and}\quad Y_{J}=\frac{1}{\sqrt{k^3}}\tilde{f}_J(\x) \quad \textmd{where}\quad \x\equiv -k\tau,$$
where $X_I$ are canonically normalized fields and $Y_I$ are non-dynamical fields which are governed by the constraint equations. Using constraints to eliminate non-dynamical quantities, and solving the equations, we can decompose the dynamical fields as
\be
X_I(\tau,k)=X_I^{G}(\tau,k)+X_I^S(\tau,k),
\ee
where $X_I^{G}(\tau,k)$ is the solution of the homogeneous equation and $X_I^{S}(\tau,k)$ is the particular part which is sourced by the other dynamical fields. Formally, we have
\be
X_I^S(\tau,k)=\frac{1}{\sqrt{k}}\int_0^{\x}G_{I}(\x,\x')S_{I}(\x')d\x',
\ee
where $G_I(\x,\x')$ and $S_{I}(\x')$ are the Green's function and source term of equation $I$ respectively. As we may expect, using the \textit{Mukhanov-Sasaki variable} 
\be\label{MS-var}
a\delta\varphi_{\Psi}\equiv a(\delta\varphi+\frac{\dot{\varphi}}{H}\Psi),
\ee
and using the constraint equations in \eqref{delta-varphi-}, we obtain the field equation of the homogeneous part of $a\delta\varphi_{\Psi}$ as
\be
(a\delta\varphi_{\Psi}^G)_{\x\x}+\bigg(1-(2+5\epsilon-3\eta)\tilde{\mathcal{H}}^2\bigg)a\delta\varphi_{\Psi}^G=0,
\ee
which is the standard equation of a single scalar field model. Imposing the standard Banch-Davis initial value for $a\delta\varphi_{\Psi}^G$, we can solve the above equation in terms of Hankel functions as
\be\label{free-axion}
a\delta\varphi_{\Psi}^G(k,\tau)=\frac{\sqrt{\pi\x}}{2\sqrt{k}}H^{(1)}_{\nu_{G}}(\x), \quad \textmd{where} \quad \nu_{G}\simeq\frac32+3\epsilon-\eta.
\ee
In order to study the contribution of the gauge field to the perturbations and determine the dynamics of the system, we will write the equations in two asymptotic limits of deep inside horizon ($\x\gg1$) and super-horizon ($\x\ll1$). The former gives us the canonically normalized fields, $\{X_I\}$s, as well as the non-dynamical fields, $\{Y_I\}$s, while the latter determines the spectral tilt and super-horizon behavior of the solutions.  The validity of our super-horizon limit analysis is crucially dependent on the stability of the scalar perturbations in the intermediate regime. That issue should be established by means of numerical study and we will address that matter in the last subsection.

\vskip 0.5 cm

\subsection{Canonically normalized fields}
At this point, after using the constraints to eliminate the non-dynamical fields in the second order action, we determine the canonically normalized fields. Setting the Banch-Davis vacuum for them, we then obtain the initial value of the rest of the variables. In the deep inside horizon limit in which $\x\gg1$, the constraint equation \eqref{dq} is \be\label{constsub}
\x\partial_{\x}(\Psi-M)+\psi\dd+\frac12\frac{\dot{\varphi}}{H}\dc=0, 
\ee
while the combination of \eqref{drho} and \eqref{A02} can be written as below
\bse\label{const--}
\begin{align}
&\frac{\dot{\varphi}}{H}\partial_{\x}\dc-\x(\Phi+\Psi)=0,\\\label{2nd-com}
&\partial_{\x}(\psi\dd+\frac12\frac{\dot{\varphi}}{H}\dc)-\x(\Psi-M)\simeq0.
\end{align}
\ese
From the combination of constraints \eqref{constsub} and \eqref{2nd-com}, up to the dominant order, we obtain
\be\label{dyn1}
\partial^2_{\x}\big(\psi\dd+\frac12\frac{\dot{\varphi}}{H}\dc\big)+\big(\psi\dd+\frac12\frac{\dot{\varphi}}{H}\dc\big)\simeq0.
\ee
Inserting \eqref{constsub} and \eqref{dyn1} into \eqref{dP} leads to $\partial_{\x}^2\Psi+\Psi=0$, which combining with \eqref{constsub} gives
\be
\partial_{\x}^2M+M=0.
\ee
Moreover, the field equation of $\dc$ \eqref{deltachi-eq} at the deep inside horizon reads as
\be\label{dc}
\partial_{\x}^2\dc+\dc=0.
\ee
The second order action up to the leading orders in $\tilde\tau$ is given by
\bea
\delta\!_{_{2}}S&=&\int a^3d^3kdt\bigg(\frac12\delta\dot\varphi^2_{\Psi}-\frac12\frac{k^2}{a^2}\delta\varphi^2_{\Psi}+\frac{3}{2}\delta\dot \psi^2-\frac{k^2}{a^2}\delta\psi^2+\frac{k^2}{a^2}\frac{(\dot M^2-\frac{k^2}{a^2}M^2)}{g^2\psi^4}+\frac{1}{2}\frac{k^4}{a^4}\frac{\tilde M^2}{H^2\psi^2}\nonumber\\
&+&\frac{k^2}{a^2}(-\frac{\tilde M}{H\psi}+\frac{2\lambda\varphi}{f}\frac{M}{g\psi^2})\delta\dot \psi+\frac{2k^2}{a^2}\frac{\lambda\varphi}{f}\frac{\dot M}{g\psi^2}\delta\psi
\bigg).
\eea
Using constraint \eqref{const--}, we can simply that to the following quadratic action
\bea
\delta\!_{_{2}}S\!\simeq\!\frac{1}{2}\int\!k^2d^3kd\tau\bigg[(a\varphi_{\Psi})_{\x}^2-(a\varphi_{\Psi})^2+2\bigg((a\dd)_{\x}^2-(a\dd)^2\bigg)
+\frac{2}{(\xi_{\psi}\psi)^2}\bigg((\x aM)_{\x}^2-(\x aM)^2\bigg)\bigg].\nonumber
\eea
The quadratic action above specifies our 3 canonically normalized (dynamical) fields as $$X_{I}=\{a\varphi_{\Psi},\sqrt{2}a\dd,-i\sqrt{2}/(\psi\xi_{\psi})\x aM\}.$$
As a result, the non-dynamical fields are
$$Y_{J}=\{a\Psi,a\Phi,a\tM\}.$$
 
Finally, imposing the standard Banch-Davis vacuum condition specifies our initial conditions as follows\footnote{It is noteworthy to mention that the above initial conditions leads to a non-vanishing scalar anisotropy 
 \be
 a^2\pi^S=\frac{iH\psi(1+\sqrt{\xi_{\psi}})}{\sqrt{k^{3}}}e^{i\x},
 \ee
 which is of the order of $\Psi$ and $\Phi$ themselves.}
\be\label{BD}
a\dc_{\Psi}=\frac{e^{i\x}}{\sqrt{2k}}, \quad a\dd=\frac{e^{i\x}}{2\sqrt{k}} \quad \textmd{and} \quad aM=\frac{i\psi\xi_{\psi}}{\x}\frac{e^{i\x}}{2\sqrt{k}}.
\ee

\vskip 0.5cm

\subsection{Long wavelength Limit and scalar spectrum}
We now turn to study the long wavelength behavior of the scalar fluctuations. The validity of our analytical calculations depends on the stability of scalar perturbations which should be established by means of numerical study. We tackle that issue in the next subsection.

At the super-horizon limit, the constraint equation \eqref{drho} has the following form
\be\label{const-sup}
V_{\varphi}\dc+6H(\dot{\Psi}+H\Phi)-\dot\varphi^2\Phi+3\big(\frac{\dot\phi^2}{a^2}
+2\frac{g^2\phi^4}{a^4}\big)\frac{\dd}{\psi}=0,
\ee
and the constraint equation \eqref{A02} is
\bea\label{AAA}
\bigg(H\dot\varphi+\frac{\lambda g\phi^2\dot{\phi}}{fa^3}\bigg)\dc+\frac{\dot{\phi}^2}{a^2}\frac{\dd}{\psi}
-2H(\dot\Psi+H\Phi)=0.~~~~~
\eea
From them, we then have
\bse\label{relation-sh}
\begin{align}
&2\bar{\rho}_{_{\rm YM}}\frac{\dd}{\psi}+\dot{\varphi}^2\Phi+\ddot{\varphi}\delta\varphi=0,\\\label{super-dq}
&V_{\varphi}\delta\varphi+6(\dot{\Psi}+H\Phi)\simeq0,
\end{align}
\ese
where the former is the combination of \eqref{const-sup} and \eqref{AAA}, while the latter is simply equation \eqref{const-sup} up to dominate orders in slow-roll.
From \eqref{dq}, \eqref{MS-var} and \eqref{super-dq}, we therefore have comoving curvature perturbation $(\mathcal{R}=\Psi-\frac{H}{\rho+P}\delta q_g)$ as 
 \be\label{super-R}
\mathcal{R}\simeq\frac{H}{\dot{\varphi}}\delta\varphi_{\Psi},
\ee
in terms of the Mukhanov-Sasaki variable.
In \eqref{free-axion}, we have the homogeneous part of $\delta\varphi_{\Psi}$, $\delta\varphi^{G}_{\Psi}$, which in super-horizon is
\be
\delta\varphi^{G}_{\Psi}(\tau,k)\simeq\frac{H}{\sqrt{2}k^{\frac32}}\x^{-(3\epsilon-\eta)}.
\ee
Moreover, the long wavelength value of the special part\footnote{It is noteworthy to mention that in our \textit{non-Abelian gauge theory}, $\delta\varphi^{S}_{\Psi}$ is coming from the contribution of linearized $F^a\tilde F_a$ to the field equation of $\dc$. In case of \textit{U(1) gauge field}, however, the linearized $F\tilde F$ vanishes and the contribution of the Abelian gauge field starts from $\delta_{2}(F\tilde F)$. In that setup, the U(1) gauge field sources the axion via inverse decay, which is now a very well studied mechanism \cite{Barnaby:2010vf, Barnaby:2011vw,Barnaby:2011qe}.
}, $\delta\varphi^{S}_{\Psi}$ can be parametrized as
\be\label{sourced-R}
\delta\varphi^{S}_{\Psi}(\tau,k)=\alpha(\xi_{\psi},\x)\frac{H}{\sqrt{2}k^{\frac32}}\x^{-(3\epsilon-\eta)},
\ee
in terms of $\alpha(\xi_{\psi},\x)$ which is a function of $\x$ and the parameter $\xi_{\psi}$. We emphasis that \eqref{sourced-R} is only a relation between wave functions, while their operators are uncorrelated. In case of stable solutions, $\alpha(\xi_{\psi},\x)$ would be a slow-varying function\footnote{Note assuming slow-roll inflation, we neglect the time variation of background parameters during the first few e-folds in which CMB fluctuations have been generated.} of $\x$, \textit{i.e.} $\alpha(\xi_{\psi},\x)\propto \x^{\mathcal{O}(\xi_{\psi},\epsilon)}$.

In order to determine $\alpha(\xi_{\psi},\x)$ and its contribution to the spectral tilt $\frac{d\ln\alpha}{d\ln k}$, we need to do numerical analysis. In the next subsection, we present the details of our numerical study of a system with $\frac{\rho_{\rm YM}}{\rho}=\epsilon^2$ and here we only summarize the final results. The homogeneous part of the comoving curvature is given as $\mathcal{R}^G=\frac{H}{\dot{\varphi}}\delta\varphi^G_{\Psi}$ which is an adiabatic mode and hence constant after horizon crossing. However, from the combination of \eqref{super-R} and \eqref{sourced-R}, we can present the special part as $\mathcal{R}^S=\alpha(\xi_{\psi},\x)\mathcal{R}^G$ (which is a functional parametrization, while the operators are uncorrelated.). Due to the prefactor $\alpha$, $\mathcal{R}^S$ can have some deviations from adiabaticity. 
 
Our scalar perturbations are stable and almost adiabatic for $\xi_{\psi}\gtrsim\sqrt{2}$ while otherwise deviates from the adiabatic solution. In particular for the parameter regime $\xi_{\psi}\gtrsim\sqrt{2}$, $\alpha(\xi_{\psi},\x)$ is almost a numerical factor of the order one ($\frac{d\ln\alpha}{d\ln k}\lesssim 10^{-3}$). 
Therefore, in the parameter regime $\xi_{\psi}\gtrsim\sqrt{2}$, we have the formal form of super-horizon power spectrum $\mathcal{R}$ as
\be\label{power-R}
P_{\mathcal{R}}=\frac{4\pi k^3}{(2\pi)^3}(\mid\!\mathcal{R}^G\!\mid^2+\mid\!\mathcal{R}^S\!\mid^2)\simeq\frac{(1+\alpha^2(\xi_{\psi}))}{2\epsilon}\bigg(\frac{H}{2\pi}\bigg)^2,
\ee
and up to the leading order in the slow-roll parameters, the spectral tilt is 
\be
n_{\mathcal{R}}-1\simeq-2(3\epsilon-\eta).
\ee
 As a result, the total comoving curvature is almost adiabatic. For smaller values of $\xi_{\psi}$, the prefactor $\alpha$ can not be considered as a numerical factor as $\frac{d\ln\alpha}{d\ln k}\gtrsim 10^{-3}$. For instance, in $\xi_{\psi}=1.2$, we have $\frac{d\ln\alpha}{d\ln k}= 10^{-2}$ and it increases rapidly as we approach smaller $\xi_{\psi}$s (see figure \ref{figScalar}).

\subsection{Stability analysis of scalar perturbations}

In the previous subsections, we analytically studied the system in two limits of sub- and super-horizon regimes. An important question that may arise and the validity of our long wavelength study tightly depends on it is the stability of scalar fluctuations in the intermediate regime. In this part, we address this important question and find the inhomogeneous solution of axion fluctuation $\delta\varphi^S$ in the presence of the gauge field. Here, we neglect the time variation of the slow-roll parameters and the metric perturbations. These slow-roll suppressed corrections may be relevant in super-horizon scales and add some small corrections to the spectral index of $\varphi^S_{\Psi}(\x)$ which we leave for future work.

\begin{figure}[h!]
\includegraphics[width=0.5\textwidth]{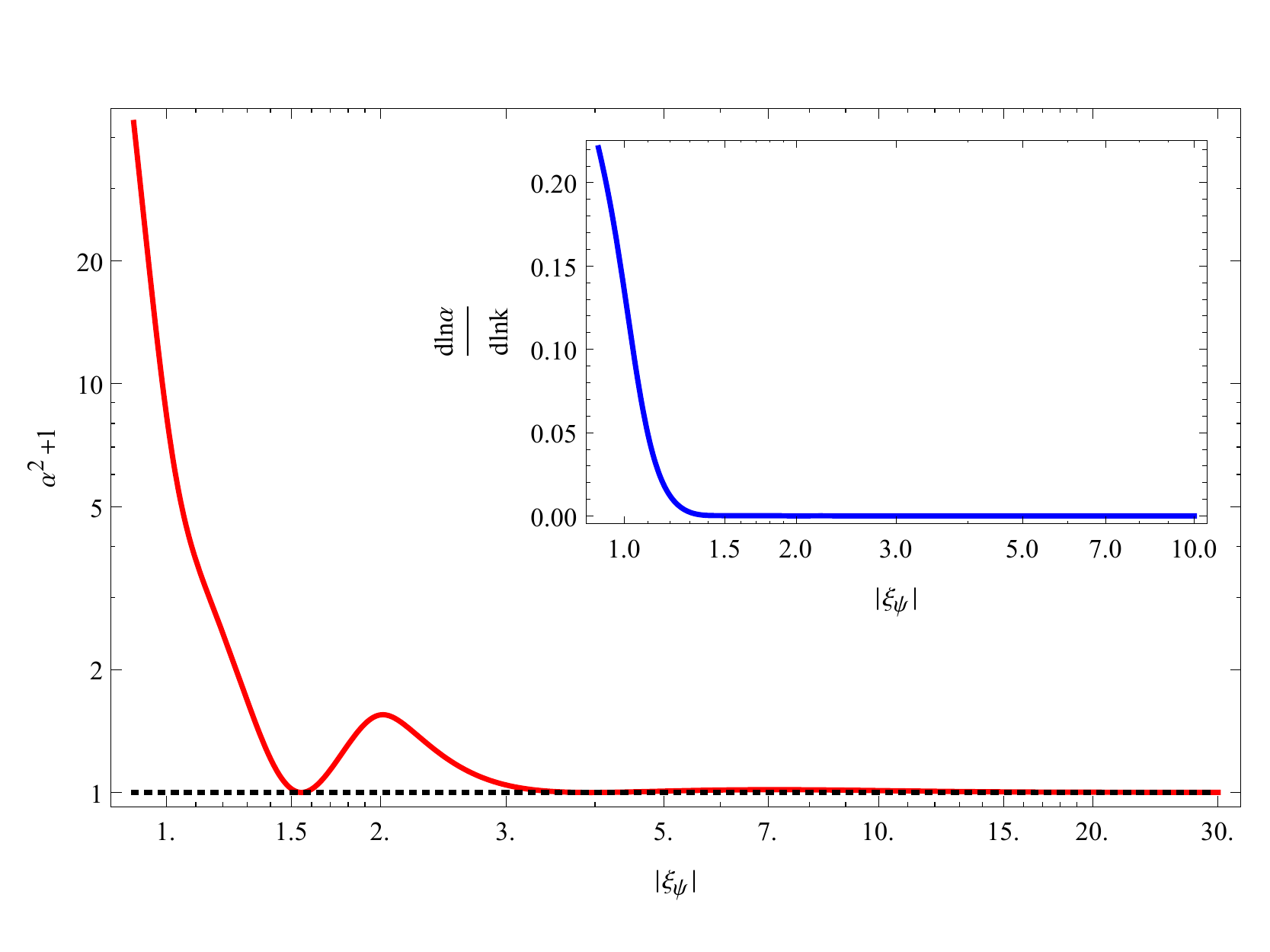}\includegraphics[width=0.53\textwidth]{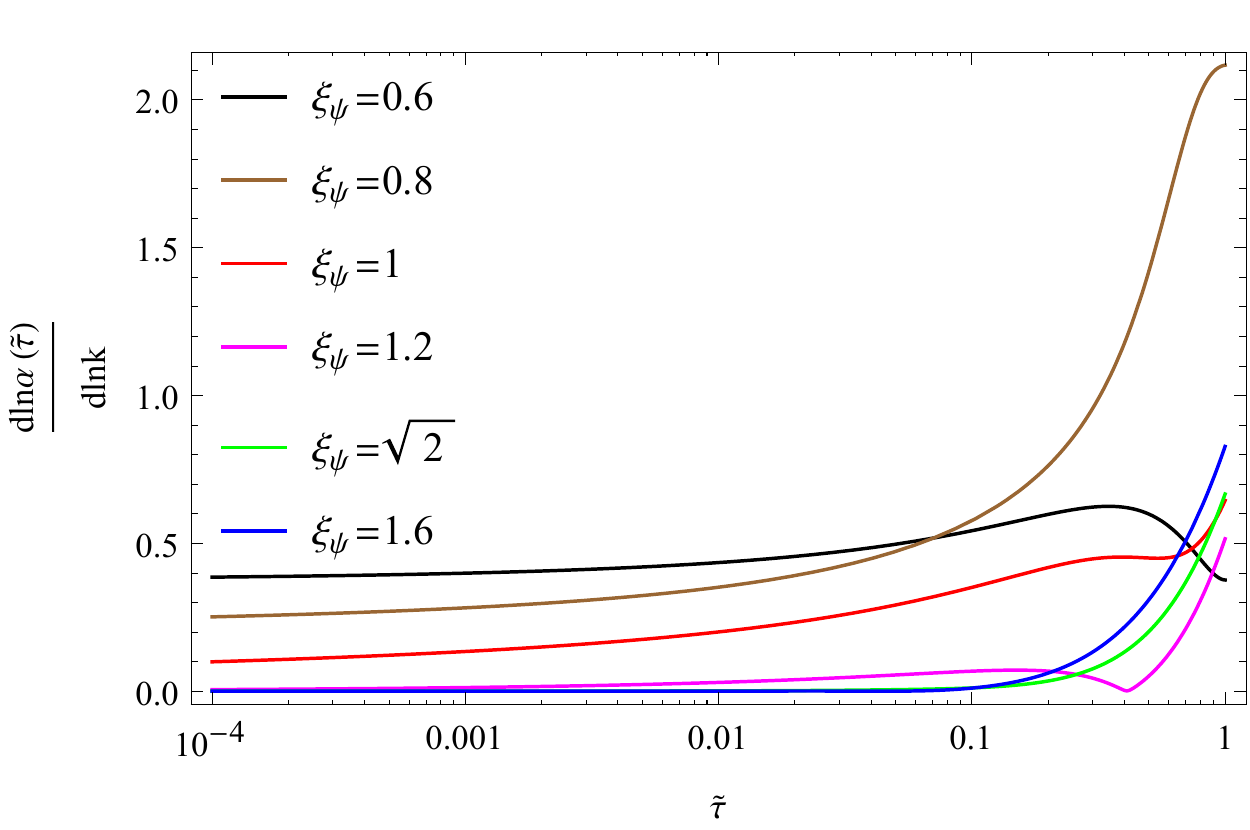}
\caption{The amplitude and the spectral index of $\delta\varphi_{\Psi}^S/\delta\varphi_{\Psi}^G$ after horizon crossing. Left panel shows $1+\alpha^2(\xi_{\psi},\x)$ and $\frac{d\ln\alpha}{d\ln k}$ at $\x=10^{-3}$ with respect to $\xi_{\psi}$. In the right panel, we present $\frac{d\ln\alpha}{d\ln k}$ vs. $\x$ for different values of $\xi_{\psi}$. The small box in the left panel shows that for $\xi_{\psi}\geqslant\sqrt{2}$, the values of $\frac{d\ln\alpha}{d\ln k}$ is less than $10^{-3}$ and therefore we can approximately consider $\alpha(\x,\xi_{\psi})$ as a numerical prefactor. However, as we go to smaller values of $\xi_{\psi}$, both of $\alpha$ and $\frac{d\ln\alpha}{d\ln k}$ increases quickly. }\label{figScalar}
\end{figure}

The special part of the axion field, $\delta\varphi^S_{\Psi}(\x)$, is sourced by the gauge field through the Chern-Simons interaction. The source term is proportional to $\frac{\lambda\psi}{f\x}$ which since $\frac{\lambda\psi}{f}\sim\sqrt{\epsilon}$, it is mostly relevant after horizon crossing. Our numerical studies show that in small scales, $\delta\varphi^S_{\Psi}$ is negligible comparing to $\delta\varphi^G_{\Psi}$, while it gradually increases as the mode approaches the horizon. After horizon crossing, for modes with $\xi_{\psi}\gtrsim\sqrt{2}$, we have $\frac{d\ln\alpha}{d\ln k}\lesssim 10^{-3}$ and therefore $\mathcal{R}^s$ is almost adiabatic (Figure \ref{figScalar}). For smaller values of $\xi_{\psi}$, on the other hand, $\delta\varphi^S_{\Psi}$ deviates from adiabatic solution. In particular, for the parameter values $\xi_{\psi}=1.2$ and $\xi_{\psi}=1$, we have $\frac{d\ln\alpha}{d\ln k}\gtrsim 10^{-2}$ and $\frac{d\ln\alpha}{d\ln k}\gtrsim 10^{-1}$ respectively. Thus, the super-horizon scalar perturbations are not adiabatic at $\xi_{\psi}\lesssim1.2$ and not even stable at $\xi_{\psi}\lesssim1$. We can also see the $\xi_{\psi}\lesssim1$ instability in the amplitude of $\alpha(\xi_{\psi},\x)$ as well. In the left panel of figure \ref{figScalar}, we present $\alpha^2+1$ vs. $\xi_{\psi}$. This quantity is almost equal to one for $\xi_{\psi}>3$, while it is around and larger than one for $\sqrt{2}<\xi_{\psi}<3$. As a result, our scalar perturbations are stable and almost adiabatic for $\xi_{\psi}\gtrsim\sqrt{2}$. In smaller values of $\xi_{\psi}$, however, it deviates from adiabatic solution and eventually becomes unstable at long wavelengths.

\section{Tensor perturbations}\label{tensor-section}

Working out the field equations of tensor fluctuations in section \ref{perturbation}, here, we turn to study the evolution of gravitational waves. The spin-2 fluctuation of the SU(2) gauge field contributes to the anisotropic stress and acts as a source term for gravitational waves.
The field equation of $h_{_{R,L}}(\tau,\textbf{k})$ in Eq. \eqref{T-gf} can be read as
\be\label{eq-h--}
\partial^2_{\x}h_{_{R,L}}+\bigg(1-\big(2-\epsilon+2\psi^2\big)\tilde{\mathcal{H}}^2\bigg)h_{_{R,L}}\simeq S^T_{_{R,L}}(\th_{_{R,L}}),
\ee
where $S^T_{_{R,L}}(\th_{_{R,L}})$ is given by the linear source term given in \eqref{T-gf}
\be
S^T_{_{R,L}}(\th_{_{R,L}})\simeq2\psi\tilde\mH\bigg(\partial_{\x}\th_{_{R,L}}+(\xi_{\psi}^2\tilde{\mH}\mp\xi_{\psi})\th_{_{R,L}}\bigg).
\ee
The solution of equation \eqref{eq-h--} can be written as 
\be\label{hR}
h_{_{R,L}}(\textbf{k},\x)=h^{G}_{_{R,L}}(\textbf{k},\x)+h^{S}_{_{R,L}}(\textbf{k},\x),
\ee
where $h^{G}$ is the homogeneous part, coming from vacuum fluctuations while $h^{S}$ is the particular part coming from the gauge field spin-2 fluctuation. We can expand $h^{G}_{R}(\textbf{k},\x)$ and $\th_{R}(\textbf{k},\x)$ as below in terms of the creation and annihilation operators\footnote{In \eqref{v-eq}, one can negligent the RHS of the equation. Therefore, gravitational waves has negligible effect on evolution of the tensor wave $\th_{R,L}$ (see equation \eqref{v-eq-app}). }
\bse\label{hG-S}
\begin{align}
h^G_{R}(\tau,\textbf{k})&=\frac{1}{\sqrt{k}}\bigg(\hat{a}^{\dag}_{\textbf{k}}h(\x)+\hat{a}_{-\textbf{k}}h^{*}(-\x)\bigg),\\\label{s-h-norm}
\th_{R}(\tau,\textbf{k})&=\frac{1}{\sqrt{k}}\bigg(\hat{b}^{\dag}_{R,\textbf{k}}\th_{R}(\x)+\hat{b}_{L,-\textbf{k}}\th_{L}^{*}(-\x)\bigg),
\end{align}
\ese
where the creation and annihilation operators satisfy the standard commutation relations
\be
[b_{\sigma,\textbf{k}},b_{\sigma,\textbf{k}'}^{\dag}]=\delta_{\sigma,\sigma'}\delta^{(3)}(\textbf{k}-\textbf{k}'), \quad [b_{\sigma,\textbf{k}},b_{\sigma,\textbf{k}'}]=[b^{\dag}_{\sigma,\textbf{k}},b^{\dag}_{\sigma,\textbf{k}'}]=0.
\ee
By definition, the left-handed polarization is given as $h_{L}(\tau,\textbf{k})=h^{*}_{R}(\tau,-\textbf{k})$.
Note that the mode functions $\frac{1}{\sqrt{k}}\th_{_{R,L}}(\x)$ and $\frac{1}{\sqrt{k}}h(\x)$ satisfy the Banch-Davis normalization, \textit{i.e.} $\frac{1}{k}\big(h(\x)h^{*'}(\x)-h^{'}(\x)h^{*}(\x)\big)=i$.
As a result, the particular part of the gravitational wave can be expanded in terms of $b_{\sigma}$ and $b^{\dag}_{\sigma}$ as
\be\label{sourced-GW}
h^S_{R}(\tau,\textbf{k})=\frac{1}{\sqrt{k}}\bigg(\hat{b}^{\dag}_{R,\textbf{k}}h^{\!^{s}}_{R}(\x)+\hat{b}_{L,-\textbf{k}}h_{L}^{\!^{s}*}(-\x)\bigg),
\ee

Note that the general solution of the tensor modes are unpolarized and is specified by one function $h(\x)$.
After imposing the Banch-Davis inertial condition to \eqref{hG-S}, we have $h$ as 
\be\label{h-hankel}
h(\x)\simeq-\sqrt{\frac{\pi\x}{2}}H^{^{(1)}}_{\nu_T}(\x) \quad \textmd{for} \quad \nu_T\simeq\frac32+\epsilon.
\ee
In order to solve the particular part of gravitational wave $h^{\!^{s}}_{_{R,L}}(\x)$, we need to determine $\th_{_{R,L}}(\x)$ in the following.

\subsection{Particular gravitational waves}

 During the slow-roll, we can neglect RHS of \eqref{v-eq}, and the field equation of $\tilde h_{_{R,L}}$ is
\be\label{v-eq-app}
\partial^2_{\x}\th_{_{R,L}}(\textbf{k},\tau)+\bigg(1\mp\frac{2(\xi+\xi_{\psi})}{\x}+\frac{2\xi\xi_{\psi}}{\x^2}\bigg)\tilde{h}_{_{R,L}}(\textbf{k},\tau)\simeq0,
\ee
in which we used the slow-roll relations \eqref{T-def}. Upon re-definitions below
\be\label{re-def}
z=-2i\x,\quad \kappa_{_{R,L}}=\mp i\big(\xi+\xi_{\psi}\big) \quad \textmd{and} \quad \mu^2=\frac14-2\xi\xi_{\psi},
\ee
we can rewrite \eqref{v-eq-app} in form of the Whittaker equation
\be
\partial^2_{z}W_{\kappa,\mu}(z)+(-\frac14+\frac{\kappa}{z}+\frac{1/4-\mu^2}{z^2})W_{\kappa,\mu}(z)=0.
\ee
The most general solutions of the above equation are Whittaker functions $W_{\kappa,\mu}(z)$ and $M_{\kappa,\mu}(z)$
\be
\th_{\sigma}(\x)=c_1 W_{\kappa\!_{\sigma},\mu}(-2i\x)+c_2 M_{\kappa\!_{\sigma},\mu}(-2i\x).
\ee
Imposing the usual Minkowski vacuum state for the gauge field's canonically normalized field $\th_{_{R,L}}$ in the asymptotic past\footnote{The $W_{\kappa,\mu}(z)$ has the following asymptotic from at the limit $\mid z\mid\rightarrow\infty$
\be
W_{\kappa,\mu}(z)\rightarrow z^{\kappa}e^{-z/2}, \quad M_{\kappa,\mu}(z)\rightarrow \Gamma(2\mu+1)\bigg(\frac{i(-1)^{\mu-\kappa}z^{\kappa}e^{-z/2}}{\Gamma({-\kappa+\mu+\frac12})}+\frac{z^{-\kappa}e^{z/2}}{\Gamma({-\kappa+\mu+\frac12})}\bigg) \quad \textmd{for} \quad \mid \arg z\mid<\frac32\pi.
\ee
Thus, the function $W_{\kappa,\mu}(-2i\x)$ represents the positive frequency solutions.}, we obtain $\th_{_{R,L}}(\x)$
\be\label{tilde-h}
\th_{\sigma}(\x)=e^{i\kappa_{\sigma}\pi/2} W_{\kappa\!_{\sigma},\mu}(-2i\x),
\ee
up to a phase factor. Moreover, the particular part of the solution is given as below
\be\label{h-sigma}
h^{\!^{s}}_{_{R,L}}(\x)=\int_{\x}^{\infty}G(\x,\x')S^T_{_{R,L}}(\x')d\x',
\ee
in which $G(\x,\x')$ is the retarded Green's function of Eqn. \eqref{eq-h--}
\bea
\label{Green}
G(\x,\x')\simeq\bigg(\frac{\x'-\x}{\x'\x}\cos(\x'-\x)-(1+\frac{1}{\x\x'})\sin(\x'-\x)\bigg)\Theta(\x'-\x),
\eea
where $\Theta(\x-\x')$ is the Heveside's delta function. 
It is useful to parametrize $h^{\!^{s}}_{_{R,L}}(\x)$ in \eqref{h-sigma} as below 
\bea\label{sourced-h--}
h^{\!^{s}}_{_{R,L}}(\x)= \bigg(\frac{\bar{\rho}_{_{\rm YM}}}{\bar{\rho}}\bigg)^{\!\frac12}\mathcal{G}_{_{R,L}}(\kappa,\mu,\x)h_{_{\rm deS}}(\x),
\eea
where $h_{_{\!\rm{deS}}}(\x)$ is the homogeneous solution of \eqref{h-hankel} in de Sitter space
\be
\frac{1}{\sqrt{k}}h_{_{\!\rm{deS}}}(\x)=\frac{1}{\sqrt{2k}}(1+\frac{i}{\x})e^{i\x},
\ee
and $\mathcal{G}_{_{R,L}}(\kappa,\mu,\x)$ is defined as
\be\label{math-G}
\mathcal{G}_{_{R,L}}(\kappa,\mu,\x)=\frac{e^{i\kappa_{_{R,L}}\!\pi/2}}{\sqrt{(1+\xi_{\psi}^2)/32}}\int^{\infty}_{\x}\frac{G(\x,\x')}{h_{_{\!\rm{deS}}}(\x)\x'}\!\biggl(\!\partial_{\x'}+(\frac{\xi_{\psi}^2}{\x'}\mp\xi_{\psi})\biggl)\!W_{\kappa_{_{R,L}},\mu}(\!-2i\x'\!)d\x'.
\ee

\begin{figure}[h!]
\includegraphics[width=0.54\textwidth]{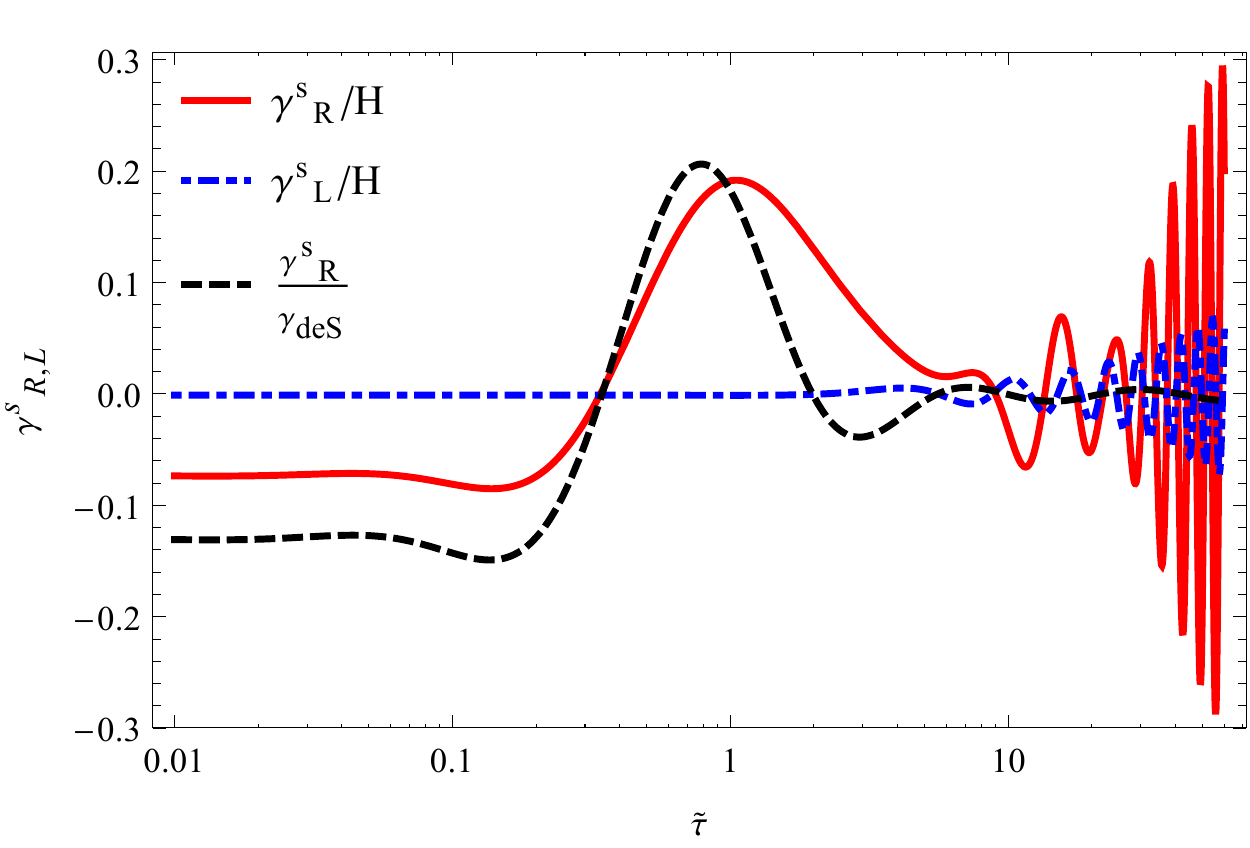}\includegraphics[width=0.5\textwidth]{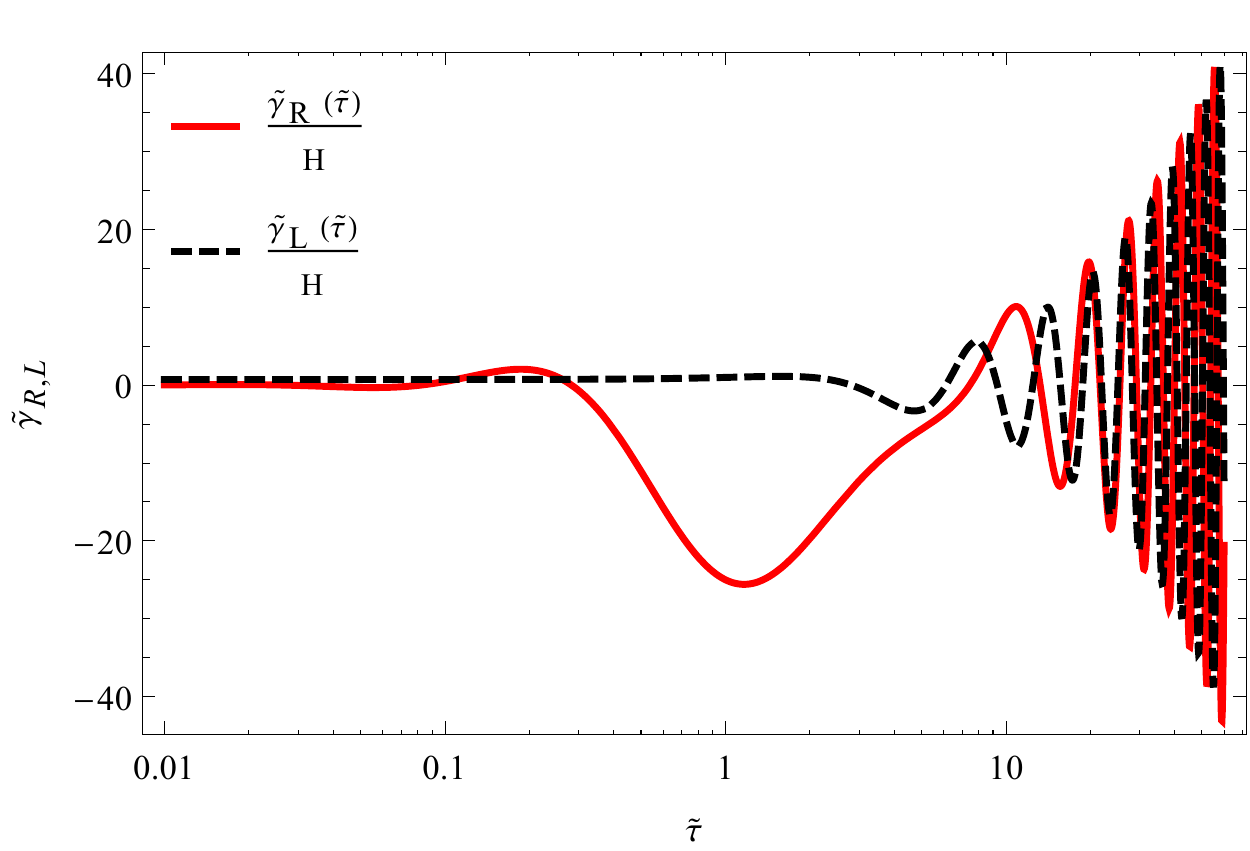}
\caption{The right panel shows $\tg{_{R,L}}$ with respect to $\x$ where the solid (red) line shows the right-handed and dashed (black) one presents the left-handed polarization. In the left panel, we plotted the particular part of gravitational waves $\gamma^s_{R}$ vs. $\x$. In this system, we choose $\rho_{_{\rm YM}}=\epsilon^2H^2$ and $\xi_{\psi}=\sqrt{3}$ and since $\psi>0$, the right-handed circular polarization is enhanced by evolution. }\label{figGW}
\end{figure}

Before analytically computing the integral \eqref{math-G} and working out the explicit form of $h^{\!^{s}}_{_{R,L}}(\x)$, here we summarize the qualitative properties of the solutions. 
As indicated by \eqref{v-eq-app}, the frequency of $\th$ gets negative for one of the polarizations for a short period before horizon crossing.
Thus, that particular polarization of $\th_{\sigma}$ experiences a short phase of tachyonic growth which eventually leads to its sharp decay after horizon crossing. The polarization with the tachyonic phase acts as an impulse function for its corresponding polarization of $h^s_{\sigma}$. That then enhances the amplitude of one of the polarizations while keeps the other polarization unchanged. In fig. \ref{figGW}, we presented the result of the numerical study of tensor fluctuations. In the following, we determine the analytic form of the particular solution of gravitational waves \eqref{h-sigma}, in the long wave length limit of the power spectrum.
\subsubsection*{$\rhd$~  super-horizon behavior of $h^{s}_{_{R,L}}$ }

\begin{figure}[h!]
\includegraphics[width=0.54\textwidth]{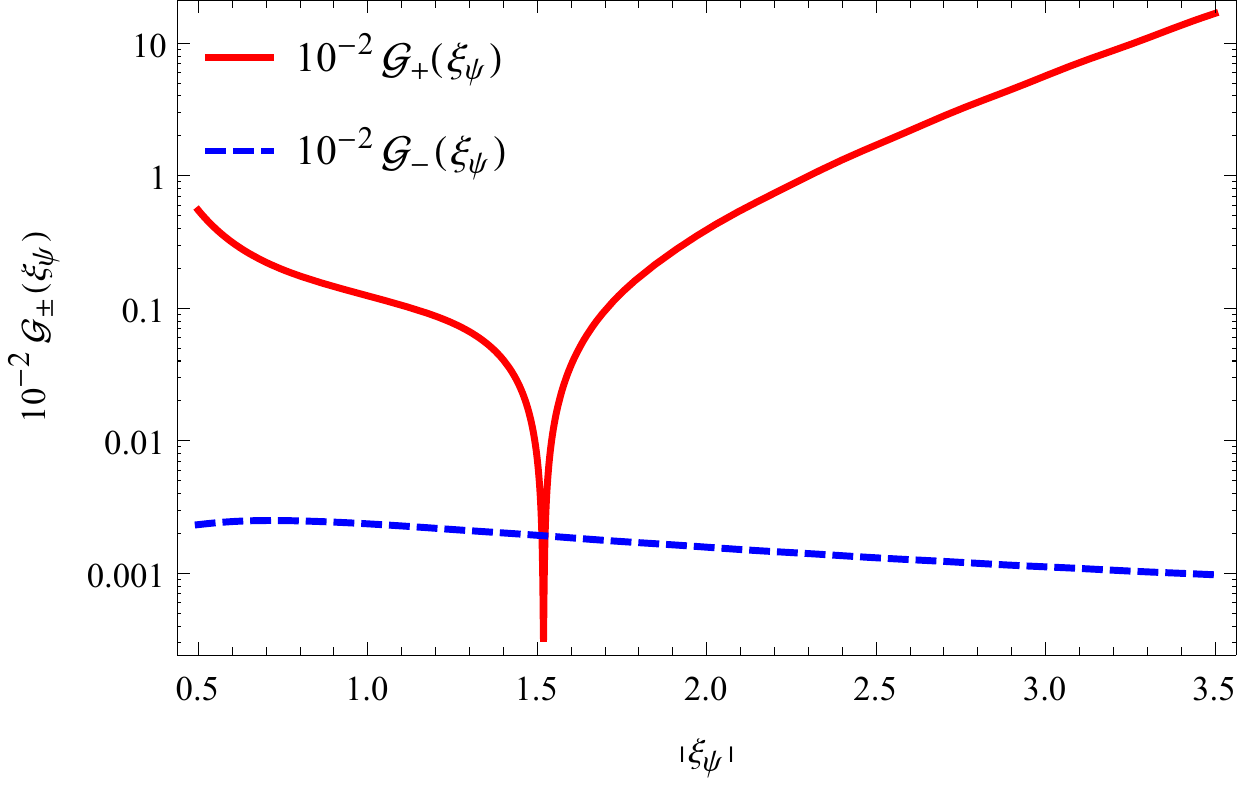}\includegraphics[width=0.54\textwidth]{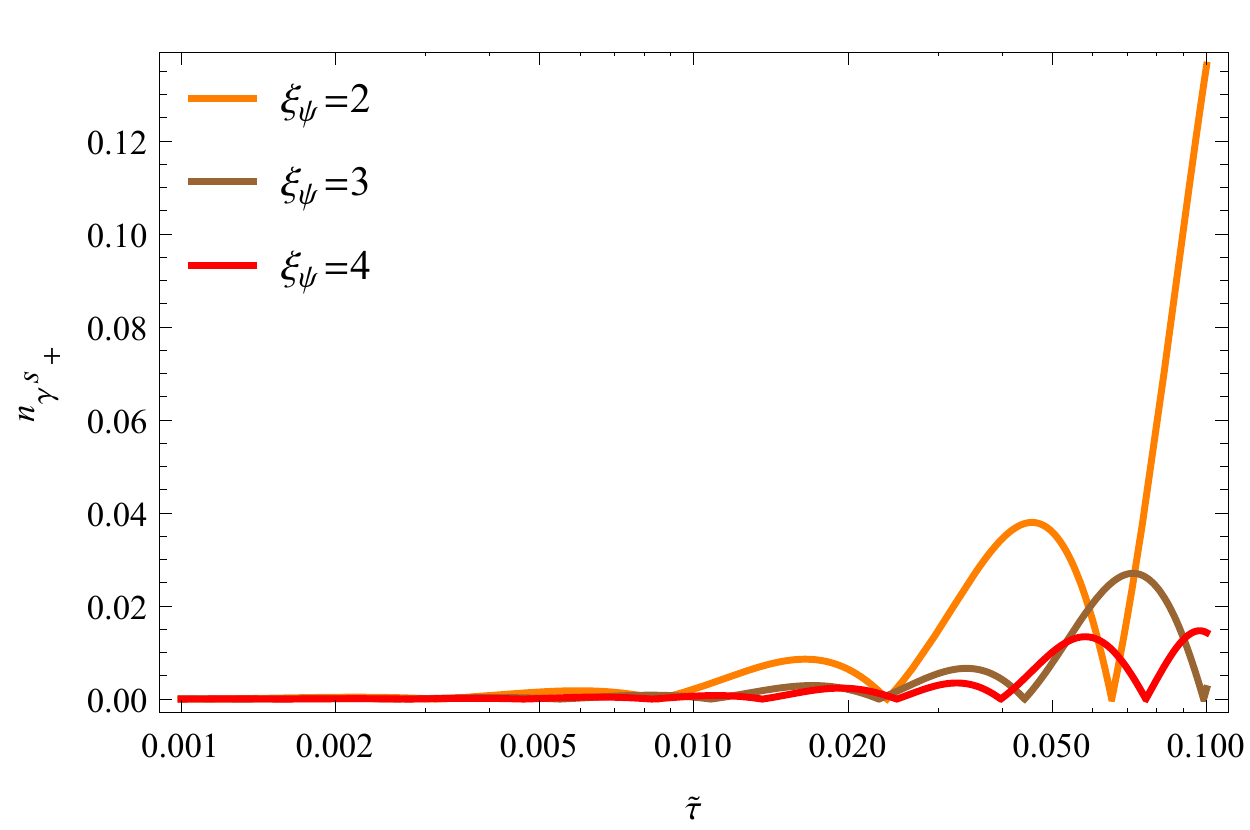}
\caption{ The left panel shows the pre-factor $\mathcal{G}_{_{\pm}}(|\xi_{\psi}|)$ with respect to $|\xi_{\psi}|$. Since $h^s_{_{\pm}}/h_{\rm deS}=\big(\frac{\rho_{\rm YM}}{\rho}\big)^{\frac12}\mathcal{G}_{_{\pm}}$ where $\big(\frac{\rho_{\rm YM}}{\rho}\big)^{\frac12}\lesssim10^{-2}$ in our model, here we presented the rescaled $\mathcal{G}_{_{\pm}}$. In the right panel, the spectral tilt of the enhanced particular mode $n_{\gamma_{+}^s}$ is illustrated with respect to $\x$ which damps like $a^{-\frac32}$.}\label{IR-Gamma}
\end{figure}

In order to study the super-horizon behavior of gravitational waves, one needs to do the Green's integral \eqref{math-G} in the limit that $\x\ll1$. We presented details of calculations in Appendix \ref{Green-int} and in the following we only report the final result. The particular solution of gravitational wave function in \eqref{sourced-h--} has the following super-horizon form
\be\label{super-h}
h^{\!^{s}}_{_{R,L}}(\x)\simeq \bigg(\frac{\bar{\rho}_{_{\rm YM}}}{\bar{\rho}}\bigg)^{\!\frac12}\mathcal{G}_{_{R,L}}(\xi_{\psi})h_{_{\!\rm{deS}}}(\x),
\ee
where the explicit form of $\mathcal{G}_{_{R,L}}$ is presented in \eqref{Int-IV}. Depending on the sign of $\psi$, the prefactor $\mathcal{G}_{_{\sigma}}$ is subleading for one of the polarization states in which $i\kappa_{\sigma}$ is negative, while it can be significant for the other one in which $i\kappa_{\sigma}>0$. We call the former integral $\mathcal{G}_{_{-}}$ and the latter one $\mathcal{G}_{_{+}}$ and have
\bse\label{super-h-}
\begin{align}
h^{\!^{s}}_{_{R,L}}(\x)&\simeq \bigg(\frac{\bar{\rho}_{_{\rm YM}}}{\bar{\rho}}\bigg)^{\!\frac12}\mathcal{G}_{_{\pm}}(\xi_{\psi})h_{_{\!\rm{deS}}}(\x)  \quad \textmd{where} \quad \psi>0,\\
h^{\!^{s}}_{_{R,L}}(\x)&\simeq \bigg(\frac{\bar{\rho}_{_{\rm YM}}}{\bar{\rho}}\bigg)^{\!\frac12}\mathcal{G}_{_{\mp}}(\xi_{\psi})h_{_{\!\rm{deS}}}(\x)  \quad \textmd{where} \quad \psi<0.
\end{align}
\ese
In the left panel of figure \ref{IR-Gamma}, we present $\mathcal{G}_{_{\pm}}$ with respect to $|\xi_{\psi}|$. Here, we rescaled  $\mathcal{G}_{_{\pm}}$ to make a more straightforward connection between the amplitude of $h^{\!^{s}}$ and $h_{_{\!\rm{deS}}}$ (in our model $\frac{\bar{\rho}_{_{\rm YM}}}{\bar{\rho}}\lesssim\epsilon^2$).

As we see, $\mathcal{G}_{_{-}}$ is always subleading and we can ignore it.
However, $\mathcal{G}_{_{+}}$ has a significant value (except around $|\xi_{\psi}|=\frac32$) and its explicit form is
\bea\label{Int-expres}
\mathcal{G}_{_{+}}(\xi_{\psi})\!&\simeq&\!e^{\frac{i\pi}{2}\kappa_{_{\!+}}}\frac{2\sqrt{(1+\xi^2_{\psi})}}{\xi^2_{\psi}} \bigg(
\frac{(i\xi_{\psi}+1)\Gamma(-\kappa_{_{\!+}})}{\Gamma(\frac12-\kappa_{_{\!+}}-\mu)\Gamma(\frac12-\kappa_{_{\!+}}+\mu)}+\frac{(i\xi_{\psi}-1)}{\Gamma(1-\kappa_{_{\!+}})}\bigg)\Gamma(\frac12-\mu)\Gamma(\frac12+\mu),\nonumber
\eea 
where $i\kappa_{+}=\frac{1+2\xi_\psi^2}{|\xi_{\psi}|}$.
As a result, the particular solution of gravitational waves are circularly polarized. In fact, depending on the sign of $\psi$, one of its polarizations gets sizeable around and after horizon crossing, while the other polarization is very small and negligible. Recalling that $\gamma^s_{\sigma}(\tau,k)=\frac{\sqrt2h^s_{\sigma}(\x)}{a}$, we have the super-horizon form for the gravitational waves ($k\tau\ll1$)
\be\label{super-hs}
\gamma^{\!^{s}}_{_{+}}(\tau,k)\simeq \bigg(\frac{\bar{\rho}_{_{\rm YM}}}{\bar{\rho}}\bigg)^{\!\frac12}\mathcal{G}_{_{+}}(\xi_{\psi})\frac{H}{k^{\frac32}} \quad  \textmd{and} \quad \gamma^{\!^{s}}_{_{-}}(\tau,k)\simeq  0.
\ee
The power spectrum of the particular solution of gravitational waves is given as
\be\label{plz-power}
P_{\gamma^{s}_{+}}=\frac{8\pi k^3}{(2\pi)^3}|\gamma^s_{+}|^2\simeq  \bigg(\frac{\bar{\rho}_{_{\rm YM}}}{\bar{\rho}}\bigg)   \mathcal{G}^2_{_{+}}(\xi_{\psi})\bigg(\frac{H}{\mpl\pi}\bigg)^2 \quad\quad  \textmd{and} \quad P_{\gamma^{s}_{-}}(\tau,k)\simeq  0,
\ee
which is circularly polarized, unlike the unpolarized vacuum fluctuation.

Due to its prefactor $(\frac{\bar{\rho}_{_{\rm YM}}}{\bar{\rho}})^\frac12  \mathcal{G}_{_{+}}(\xi_{\psi})$ in \eqref{super-hs}, $\gamma^s_{+}$ does not \textit{exactly} freeze out after horizon crossing, but it evolves slowly as 
\be
\frac{d\ln\gamma^s_{+}(\tau,k)}{d\ln\tau}=-\vartheta-(\epsilon+\vartheta)\frac{d\ln(\sqrt{(1+\xi^2_{\psi})}\mathcal{G}_{_{+}})}{d\ln\xi_{\psi}}.
\ee
%here we used the fact that $\frac{\dot{\xi_{\psi}}}{H\xi_{\psi}}=\vartheta+\epsilon$.
and therefore is slightly deviates from the adiabatic solution, $\frac{d\ln\gamma^s_{+}(\tau,k)}{d\ln\tau}=\mathcal{O}(\epsilon)$. The spectral tilt of $\gamma^{\!^{s}}_{_{+}}$ has a rather complicated behavior
which is presented in the right panel of figure \ref{IR-Gamma}. It has damped oscillations which decays as $a^{-\frac32}$ at large scales and fades away. %That is one of the peculiar features of the tensor modes in our model.

\subsection{Modified Lyth bound and tensor spectrum}

\begin{figure}[h!]
\begin{center}
\includegraphics[width=0.55\textwidth]{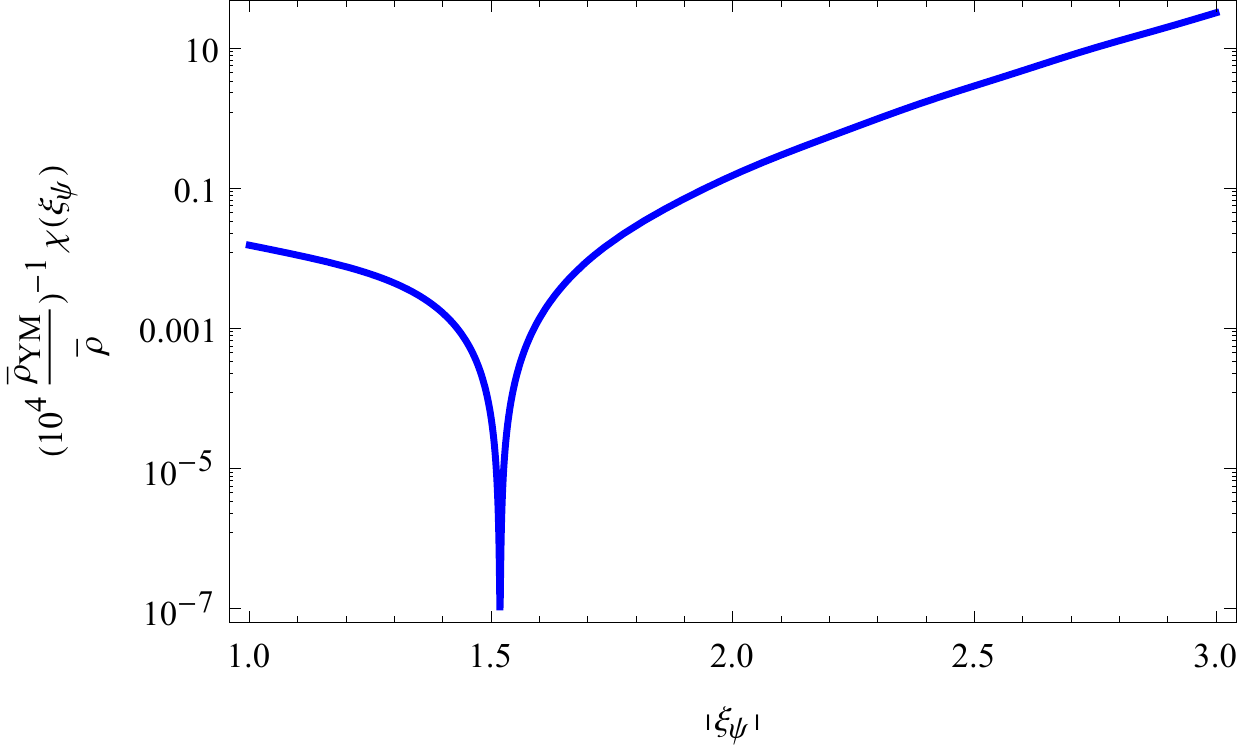}\includegraphics[width=0.5\textwidth]{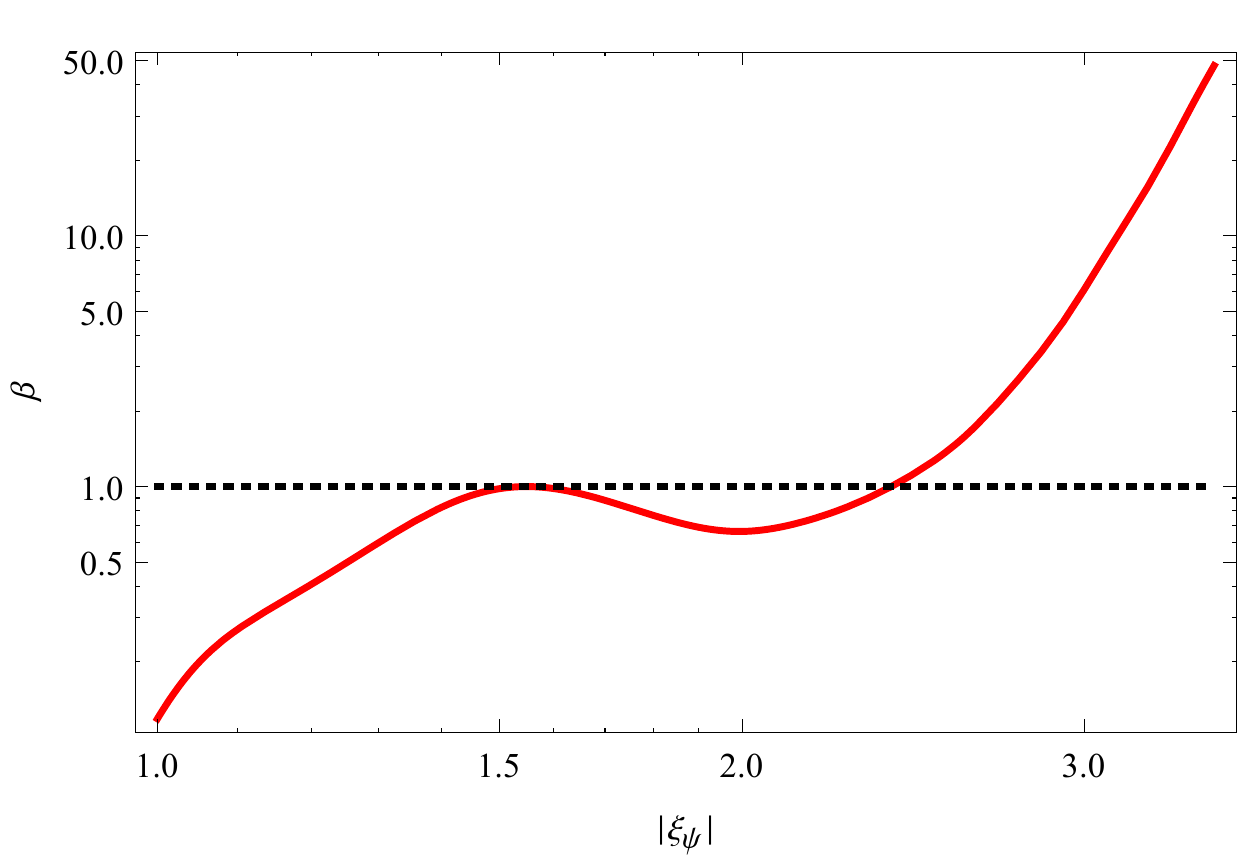}
\caption{The chirality parameter $\chi$ and $\beta$ with respect to $\x$ for a system with $\frac{\bar{\rho}_{_{\rm YM}}}{\bar{\rho}}=\epsilon^2$.}\label{r}
\end{center}
\end{figure}

Given the fact that $h^{G}$ and $h^{S}$ are uncorrelated and working out \eqref{h-hankel} and \eqref{super-h-}, we obtain the power spectrum of gravitational waves as
\be\label{power-T}
P_{_{T}}\simeq \bigg(2+\frac{\bar{\rho}_{_{\rm YM}}}{\bar{\rho}}\mathcal{G}^2_{_{+}}(\xi_{\psi})\bigg)\big(\frac{H}{\pi\mpl}\big)^2 .
\ee
In fact, the gauge field's tensor fluctuations modified the gravitational waves power spectrum proportional to $\frac{\bar{\rho}_{_{\rm YM}}}{\bar{\rho}}$ and a function of $\xi_{\psi}$. However, the tensor spectral tilt of vacuum fluctuations is the same as the standard one 
\be
n_T=-2\epsilon,
\ee
 One of the polarization states of $\gamma_{ij}$ has the power spectrum of graviton vacuum fluctuations, $P_{_{vac}}(\x)\simeq\big(\frac{H}{\pi\mpl}\big)^2$, while the other is enhanced by the gauge field ( see equation \eqref{plz-power}). We can parametrize the chirality of CMB power spectrum by the dimensionless parameter 
\be
\chi\equiv\frac{P_{_{R}}-P_{_{L}}}{P_{_{vac}}}=s\mathcal{G}^2_{_{+}}(\xi_{\psi})\frac{\bar{\rho}_{_{\rm YM}}}{\bar{\rho}}, \quad \textmd{where} \quad
s=\textmd{sign}(\psi).
\ee
In the left panel of figure \ref{r}, we present $\chi$ with respect to $\xi_{\psi}$. As we see, it is negligible if $\mid\!\xi_{\psi}\!\mid\lesssim\frac32$, however it increases monotonously for $\mid\!\xi_{\psi}\!\mid>\frac32$.

 The other important observational quantity is tensor to scaler ratio $r$ and using \eqref{power-R} and \eqref{power-T}, the prediction of our models is
\be
r=16\epsilon\beta \quad \textmd{where} \quad \beta\equiv\bigg(\frac{1+\frac{\bar{\rho}_{_{\rm YM}}}{2\bar{\rho}}\mathcal{G}^2_{_{+}}(\xi_{\psi})}{1+\alpha^2(\xi_{\psi})}\bigg). 
\ee
The right panel of figure \ref{r}, shows $\beta$ for $\frac{\bar{\rho}_{_{\rm YM}}}{\bar{\rho}}\sim\epsilon^2$ with respect to $\xi_{\psi}$.
As we see here, $\beta$ increases by $\mid\!\xi_{\psi}\!\mid$ for $\mid\!\xi_{\psi}\!\mid<\frac32$ and $\mid\!\xi_{\psi}\!\mid>2.5$. $\beta$ is less than one for $\mid\!\xi_{\psi}\!\mid<2.5$, while is more than one and increases sharply by $\mid\!\xi_{\psi}\!\mid$ otherwise.

Lyth (1997) noted that for standard single scalar slow-roll inflation, we can relate the change in the inflaton during inflation, $\Delta\varphi$, to the tensor to scalar ratio and the number of e-folds $N$, as $\Delta\varphi\sim\mpl N\sqrt{\frac{r}{8}}$ \cite{Lyth:1996im}. In our setup, slow-roll inflation is driven by the axion potential. The SU(2) gauge field is negligible on the background level, however, it has a significant contribution on the scalar and tenor perturbations. Therefore, our model satisfies in the following modified version of Lyth bound
\be
\Delta\varphi\sim\mpl N_{CMB}\sqrt{\frac{r}{8\beta}},
\ee
which relates the axion excursion and $r$.

\subsection{Generic features of tensor fluctuations}

In this subsection, we summarize the generic features of the tensor perturbations in our model.

\begin{itemize}
\item{We have two tensor fluctuations $\gamma_{ij}$ and $\tg_{ij}$ which are coupled to each other. The former is the \textit{gravitational wave} coming form the perturbed metric while the latter is the spin-2 fluctuations of the perturbed SU(2) gauge field, \textit{tensor waves}.}
\item{The sound speed of both  $\gamma_{ij}$ and $\tg_{ij}$ are equal to one.}
\item{Our system is diagonalized in terms of the circular polarizations. In particular, there are parity odd terms in the perturbed action which have different signs for the right- and left-handed polarization states.}
\item{Due to its parity odd interactions, one of the polarization states of $\tg_{ij}$ experiences a short period of tachyonic growth before horizon crossing, around $\frac{k}{aH}=2(\xi_{\psi}+\xi)\sim\mathcal{O}(1)$. Shortly after that, however, it starts to decay and fade away.}
\item{The effective mass of $\tg_{ij}$ is equal to $2(1+\xi_{\psi}^2)H^2$ which leads to decay of its \textit{both} polarizations after horizon crossing.}
\item{$\tg_{ij}$ contributes to the anisotropic stress $\pi^T_{ij}$ and acts as a source term for the gravitational waves. Thus we can decompose $\gamma_{ij}$ into its vacuum fluctuations, $\gamma^{G}_{ij}$, and the particular solution $\gamma^{S}_{ij}$ which is sourced by the SU(2) gauge field.}
\item{Our vacuum solutions $\gamma^{G}_{ij}$ is \textit{unpolarized} and has the same amplitude as the standard vacuum gravitational waves in the scalar inflationary models. }
\item{The particluar part of gravitational waves, $\gamma^{S}_{ij}$, is circularly polarized. Both of its polarization states are subdominate inside the horizon. However, one of its polarizations $\gamma^s_{+}$, is enhanced around horizon crossing while the other one, $\gamma^s_{-}$, is always negligible. }
\item{If $\psi$ is positive/negative, the right-/left-handed polarization of $\gamma^{S}_{\sigma}$ would get  enhanced by its corresponding $\tg_{\sigma}$ field around the horizon crossing. Therefore, the total tensor power spectrum is modified by a factor proportional to $\frac{\bar\rho_{\rm YM}}{\bar{\rho}}$. Since this modification is only on one polarization state, that generates a chirality equal to $\frac{P_{_{R}}-P_{_{L}}}{P_{_{vac}}}=\textmd{sign}(\psi)\mathcal{G}^2_{_{+}}(\xi_{\psi})\frac{\bar{\rho}_{_{\rm YM}}}{\bar{\rho}}$.
As a result, our setup predicts non-vanishing parity odd CMB correlations, $\ev{TB}$ and $\ev{EB}$.}
\item{Because of the spin-2 fluctuations of the SU(2) gauge field, the total power spectrum is enhanced with respect to the vacuum fluctuations, \textit{i.e.} $P_{\rm T}=(1+\frac{\bar{\rho}_{\rm YM}}{2\bar{\rho}}\mathcal{G}_{+}^2)P_{\rm T}^{vac}$. That breaks the direct relation between the power spectrum of the gravitational waves and the scale of inflation.}
\item{The tensor to scalar ratio and the Lyth bound are also modified. In particular, the tensor to scalar ratio and the axion excursion are now given as $r=16\beta\epsilon$ and $\Delta\varphi\sim\mpl N\sqrt{\frac{r}{8\beta}}$ where $\beta$ is presented in figure \ref{r}.}
\end{itemize}

\section{Discussion}\label{conclusion}

In this paper, we have studied the very well-motivated axion inflation models in the presence of an SU(2) gauge field with a small (but non-vanishing) vev. We found that although the gauge field has a small energy density $\rho_{_{\rm YM}}\lesssim\epsilon^2H^2$, yet it leads to a rich phenomenology and new observables in the CMB anisotropy. The inflaton field is the axion $\varphi$ which for the sake of generality has an arbitrary potential. Thanks to the non-Abelian nature of the gauge field, it can have a homogeneous and isotropic solution and therefore a background energy density. Moreover, the Chern-Simons interaction ($\frac{\lambda\varphi}{4f}\tilde F^aF_a$) breaks the conformal invariance of the gauge field and prevents its decay during inflation.  As the axion rolls down its potential, $\dot{\varphi}/H$ increases and part of the energy of the axion gradually injects to the gauge field, hence $\rho_{_{\rm YM}}$ slowly increases during inflation. After the end of inflation, on the other hand, $\dot{\varphi}$ starts oscillating around the minimum of the potential and the gauge field acts like a dark radiation, $\rho_{_{\rm YM}}\propto a^{-4}$. Therefore, in this scenario, inflation ends in a self-interacting dark radiation dominated Universe which may have interesting features for the (pre)reheating era. Moreover, the interaction $\varphi F^a\tilde F_a$ provides a natural decay channel for the inflaton during (pre)reheating which is beyond the scope of this paper.  The slow-roll dynamics of the gauge field requires that $\frac{\lambda}{f}\sim\frac{\mathcal{O}(10)}{\mpl}$. Since large coupling is hard to achieve in a controlled string compactification \cite{Baumann:2014nda}, here we are interested in small values of $\lambda$.

The SU(2) gauge field has a negligible contribution to the inflation dynamics, however, it leaves notable features on the cosmic perturbations.  Its fluctuations can be decomposed into scalar, vector and tensor modes. The scalar perturbations are modified by the gauge field at large scales while the vector fluctuations are still damping and unimportant.  The scalar perturbations are stable and almost adiabatic for $\xi_{\psi}\gtrsim\sqrt{2}$ while otherwise deviates from the adiabatic solution. Moreover, in the parameter regime $\xi_{\psi}\lesssim1$, the scalar perturbation is unstable.  Tensor perturbations are also modified by the gauge field. In particular, the SU(2) gauge field has a spin-2 perturbation which is coupled to the primordial gravitational waves. This new tensor fluctuation explicitly breaks the parity between the left- and right-handed polarization states. Our gravitational waves are the standard vacuum fluctuations plus the  particular solution coming from the spin-2 fluctuations of the gauge field. The former has the standard power spectrum $P_T^{\rm vac}=2\big(\frac{H}{\pi\mpl}\big)^2$ while the latter has a polarized power, proportional to the background energy density of the gauge field and a prefactor function of $\xi_{\psi}$, $P^{+}_T=\frac{\bar{\rho}_{_{\rm YM}}}{\bar{\rho}}\mathcal{G}^2_{_{+}}(\xi_{\psi})\big(\frac{H}{\pi\mpl}\big)^2$. $P^{+}_T$ is the circularly polarized part of the gravity waves power spectrum and quantifies the amounts of chirality in the super-horizon power spectrum. That results in parity odd CMB correlations between E and B-modes and T and B-models.  In the parameter regime $\sqrt{2}<\xi_{\psi}<3$, the gauge field generates  simultaneously a detectable chiral gravitational wave signal with negligible contribution to the scalar fluctuations, in agreement with the current CMB observations. Hence the axion excursion satisfies in a modified version of the Lyth bound and scale of inflation is not directly related to the tensor power spectrum.

We emphasise that the perturbed SU(2) gauge field is linearly coupled to the gravitational wave. This is in contrast to the case of U(1) gauge field in which the Abelian gauge field quanta is mixed to the gravitational waves at the nonlinear level through $\varphi F\tilde F$. In  that construction of axion driven inflations, the U(1) gauge field quanta are also coupled to the curvature and generates large amounts of non-Gaussianity. Therefore, the resulting gravity wave signal is correlated to the large scale non-Gaussianity \cite{Barnaby:2011qe, Barnaby:2011vw}. In the non-Abelian case, however, the mixing between the gauge field and perturbations in the scalar and tensor sectors i) are coming from different fluctuations and ii) at the linear order. Hence, the enhancement of gravitational wave and the modification in the scalar perturbations are uncorrelated. Given the mixing between the inflaton field and the SU(2) gauge field, perhaps the most important question that is left to answer is the non-Gaussianity of this scenario, which we postpone for future work. 

One of the interesting and robust features of  this setup is the generation of intrinsic chiral gravity waves which makes it distinguishable from the unpolarized vacuum fluctuations. Interestingly, the spin-2 fluctuations of the SU(2) gauge field provide a source of CP violation during inflation. Inspiring by the gravitational leptogenesis scenario introduced in \cite{Alexander:2004us}, one may explore the possibility of the lepton production during inflation. In \cite{Maleknejad:2016dci}, using the gravitational anomaly in the standard model of particle physics, we studied that possibility. 
We found that this setup can serve as a leptogenesis mechanism during inflation and explain the observed baryon asymmetry in the Universe.

\section*{\small Acknowledgment}

It is a pleasure to thank Peter Adshead, Lorenzo Bordin, Paolo Creminelli, Tomohiro Fujita and Marco Peloso for helpful discussion. I am grateful to the hospitality of Stanford University where this  work has been initiated and the Galileo Galilei Institute for theoretical physics (GGI) and INFN during its completion. I acknowledge support from Allameh Tabatabaii grant of Boniad Melli Nokhbegan Iran.

\appendix
\section{Geometry of gauge invariant combinations}\label{gauge-invariant}

The perturbed FRW metric can be parametrized as
\be\label{metric-pert-app}%
ds^2=-(1+2A)dt^2+2a(\partial_iB+V_i)dx^idt+a^2\left((1-2C)\delta_{ij}+2\partial_{ij}E+2\partial_{(i}W_{j)}+\gamma_{ij}\right)dx^idx^j\,,
\ee
where $A,\ B,\ C$ and $E$ parametrize scalar perturbations, $V_i,\ W_i$ are vector perturbations and $\gamma_{ij}$ is the symmetric, traceless and divergence-free tensor mode. We can also define the tetrad field $e^a_{\mu}$
\be
g_{\mu\nu}=\eta_{\alpha\beta}e^{\alpha}_{~\mu}e^{\beta}_{~\nu},
\ee
where $\eta_{\alpha,\beta}$ is the Minkowski metric and $\alpha, \beta$ runs from $0$ to $3$. One can choose the background tetrads as
\be
\bar{e}^{0}_{~\mu}=n_{\mu} \quad \textmd{and} \quad \bar{e}^{a}_{~\mu}=a(t)\delta^a_{\mu},
\ee
where $n^{\mu}=(1,0,0,0)$ is the 4-velocity of the comoving observer. From the perturbed metric we can set
\be
\delta{e}^{0}_{~\mu}=\delta n_{\mu}-\delta e^a_{~0}\bar{e}_{a\mu} \quad \textmd{and} \quad e^{a}_{i}=\delta g_{ij}\bar{e}^{aj},
\ee
where $\delta n_{\mu}=(-A,a\partial_iB+aV_{i})$. For later convenience, we choose the perturbed tetrad fields as
\bse \label{tetrad-pert-Appn}
\begin{align}
\delta e^a_{~i}&=a\bigg(-C\delta^{a}_{i}+\delta^{aj}(\partial_{ij}E+\partial_{(i}W_{j)}+\frac12\gamma_{ij})\bigg), \quad \delta e^0_{i}=-a^2\partial_i(\dot E-\frac{B}{a}),\nonumber\\
\delta e^a_{~0}&=\delta^{aj}(a\partial_j\dot{E}+V_j), \quad \quad \quad \delta e^0_{~0}=-A.
\end{align}
\ese
The axion and SU(2) gauge field are also perturbed around their homogeneous and isotropic background configurations (Eqn. \eqref{field-perturb}) as
\be\label{field-appdx}
\varphi(t,\textbf{x})=\varphi(t)+\delta\tilde{\varphi}(t,\textbf{x})\quad \textmd{and}\quad A^a_{~\mu}(t,\textbf{x})= \psi(t)\bar{e}^a_{\mu}(t)+\delta A^a_{~\mu}(t,\textbf{x}),
\ee%
where $\delta A^a_{~\mu}$ involves $3\times4$ components. Therefore, the 13 field perturbations together with the 10 components of the perturbed metric, add up to 23 degrees of freedom. Due to the gauge transformations, not all of that metric and field perturbations are gauge invariant.
In particular, we have two types of gauge freedoms: we call them ``$x^\mu$-gauge'' and ``$A^a$-gauge''.
\begin{itemize}
\item{\textit{$x^\mu$-gauge} are the space-time gauge transformations
\be\label{coord}
x^\mu\mapsto x^\mu+\xi^{\mu}(t,\textbf{x}) %\quad \textmd{where} \quad \xi^{\mu}=\left\{\begin{array}{ll} \partial_i\delta x+\delta x_i^V \quad &\mu=i,\\
%\delta t \quad &\mu=0,
%\end{array}\right.
\ee
%where $\delta x$ and $\delta t$ are two scalars and $\delta x^V_i$ is a transverse vector.
which acts on the perturbed metric and fields as follows
\bse
\begin{align}
\delta g_{\mu\nu}&\mapsto \delta g_{\mu\nu}-\mathcal{L}_{\xi}\bar{g}_{\mu\nu}=\delta g_{\mu\nu}-\delta t\dot{\bar{g}}_{\mu\nu}-2\bar{g}_{\lambda(\nu}\partial_{\mu)}\delta x^{\lambda},\\
\delta\tilde{\varphi}&\mapsto \delta\tilde{\varphi}-\dot\varphi\delta t,\\\label{gauge-trans}
\delta A^a_{~\mu}&\mapsto \delta A^a_{~\mu}-\dot{\bar{A}}^a_{~\mu}\delta t-\bar{A}^a_{~\nu}\partial_{\mu}\xi^{\nu}=\delta A^a_{~\mu}-\dot{\psi}\bar{e}^a_{~\mu}\delta t-\psi\mathcal{L}_{\xi}\bar{e}^a_{~\mu},
\end{align}
\ese
where $\mathcal{L}_{\xi}$ is the Lie derivative with respect to $\xi^{\mu}$.

%The gauge invariant combination constructed from $\delta\tilde\varphi$ is $\dc=\delta\tilde\varphi-\dot\varphi a^2(\dot{E}-\frac{B}{a})$.
As we see in \eqref{gauge-trans}, due to its vector nature, the perturbed gauge field changes under the action of the space-time gauge transformations. Thus, it is useful to decompose $\delta A^a_{~\mu}$ as
$$\delta A^a_{~\mu}=\delta_{_{x}} A^a_{~\mu}+\delta\!_{_{g\!f}} A^a_{~\mu},$$
in which $\delta_{_{x}} A^a_{~\mu}$ is the \textit{induced space-time} transformations on the gauge field, and $\delta\!_{_{g\!f}} A^a_{~\mu}$ is the \textit{genuine gauge field} fluctuations which is invariant under the action of $x^\mu$-gauge. As one may expect from \eqref{field-appdx}, equation \eqref{gauge-trans} then specifies $\delta_{_{x}} A^a_{~\mu}$ as
\be
\delta_{_{x}} A^a_{~\mu}=\psi\delta e^a_{~\mu}.
\ee}
\item{\textit{$A^a$-gauge} is the infinitesimal \textit{internal gauge field} transformation which acts on the gauge field as
\be \label{gua}
\delta\!_{_{g\!f}} A^a_{~\mu}\mapsto \delta\!_{_{g\!f}} A^a_{~\mu}+\frac{1}{g}D_\mu\lambda^a, 
\ee
where $D_{\mu}=\partial_{\mu}+igA_{\mu}$ is the covariant derivative. The gauge transformation parameter $\lambda^a(t,\textbf{x})$ can be decomposed as
$$\lambda^a=\delta^{ai}\partial_i\lambda+\delta^a_i\lambda^{~i}_{V},$$
in which $\lambda$ is the scalar and $\lambda^V_i$ is the divergence-free vector parts.}

\end{itemize}

Thus, 12 components of $\delta A^a_{~\mu}(t,\textbf{x})$ can be decomposed as (Eqn. \eqref{gauge-field-pert})
\bse \label{gauge-field-pert-Appn}
\begin{align}
\delta A^a_{~i}&=a\delta^a_i \delta\psi+\delta^{aj}\big(\partial_{ij}\tilde{Z}+\partial_i v_j+a\tg_{ij}\big)+a\psi\epsilon^{a~j}_{~i}\big(g\partial_{j}(Z-\tilde{Z})+w_j\big)+\psi \delta e^a_{~i},\nonumber\\
\delta A^a_{~0}&=\delta^{k}_a\partial_kY+\delta_a^j u_j+\psi \delta e^a_{~0},
\end{align}
\ese
in which $\{\dd, Y, \tilde Z, Z, u_i, v_i, w_i,\th_{ij}\}$ are the genius gauge fluctuations and therefore invariant under the infinitesimal space-time gauge transformations \cite{Maleknejad:2012fw}. The explicit form of $\delta e^{\alpha}_{~\mu}$ is presented in \eqref{tetrad-pert-Appn}. 

 Now we are ready to construct the gauge invariant combinations of each sector.

\vskip 0.5cm
$\triangleright$  \textit{Scalar modes}

In the scalar sector of the perturbations, $A$, $B$, $C$, $E$ are coming from the perturbed metric and,
$\dd$, $Y$, $Z$ and $\tilde{Z}$ from the perturbations of the gauge field.
Under the action of the transformation \eqref{coord} ($\xi^0=\delta t,\xi^i=\partial_i\delta x$), the scalar fluctuations of the metric transform as
\be\label{met-x}
\begin{split}
&A \mapsto A-\dot{\delta t}\,,\qquad \qquad \quad
C \mapsto C+H\delta t\,,\\
&B \mapsto B+\frac{\delta t}{a}-a\dot{\delta x}\,,\qquad
E \mapsto E-\delta x\,,
\end{split}
\ee
and $\delta\tilde\varphi$ changes as
\be
\delta\tilde\varphi \mapsto \delta\tilde\varphi-\dot\varphi\delta t.
\ee
By definition, the genuine gauge scalars $\{\dd, Y, Z, \tilde{Z}\}$ are invariant under the $x^\mu$-gauge transformations.
On the other hand, under the action of the internal gauge field transformation of the form \eqref{gua}, the gauge field perturbations transform as
\be\label{A-g}
\begin{split}
\dd \mapsto \dd\,& ,\qquad
Y \mapsto Y-\frac{1}{g}\dot{\lambda}\,,\\
Z\mapsto Z\,&,\qquad
\tilde{Z}\rightarrow \tilde{Z}-\frac{1}{g}\lambda\,.
\end{split}
\ee
From the combination of \eqref{met-x} and \eqref{A-g}, we then can construct six independent gauge-invariant combinations; the standard Bardeen potentials from the metric perturbations 
\bse
\begin{align}
\Psi=&C+a^2H(\dot{E}-\frac{B}{a})\,,\\
\Phi=&A-\frac{d}{dt}\left(a^2(\dot{E}-\frac{B}{a})\right)\,,
\end{align}
\ese
as well as the matter combinations
\bse
\begin{align}
\dc=&\delta\tilde\varphi-\dot\varphi a^2(\dot{E}-\frac{B}{a}), \qquad \dd=\dd,\\
 M=&\frac{g^2\phi^3}{a^2}Z, \qquad \qquad \qquad \quad \tilde{M}=\dot\phi(\dot{\tilde{Z}}-Y)\,,
\end{align}
\ese
which are coming from the axion and gauge field fluctuations.

\vskip 0.5cm
$\triangleright$  \textit{Vector modes}

In the vector sector, we have $V_i$, $W_i$, $u_i$, $v_i$ and $w_i$ which
under the action of an infinitesimal ``vector'' coordinate transformation \eqref{coord} ($\xi^0=0,\xi^i=\delta x_{V}^i$), transform as
\be
\begin{split}
V_i\mapsto V_i-a\delta\dot{x}_V^i\,&,\qquad W_i \mapsto W_i-\delta x_V^i\,.\\
\end{split}
\ee
$u_i$ and $v_i$ remain invariant under the coordinate transformations, however, under the infinitesimal gauge transformation \eqref{gua}, they change as
\be
u_i \mapsto u_i-\frac1g\dot{\lambda}_V^i\,,\quad
v_i \mapsto v_i-\frac1g\lambda_V^i\,,\quad
w_i \mapsto w_i+\lambda_V^i\,.
\ee
The metric fluctuations $V_i$ and $W_i$ obviously remain unchanged under \eqref{gua}.

We can construct three gauge invariant divergence-free vector perturbations, one from the metric fluctuation
\be
\mathcal{Z}_i=a\dot{W}_i-V_i\,,
\ee
and two from our genuine gauge field perturbations
\bea
\mathcal{U}_i=\frac1g\dot{w}_i+u_i\,,\quad \textmd{and} \quad \mathcal{V}_i=\frac1g w_i+v_i\,.
\eea

\vskip 0.5cm
$\triangleright$  \textit{Tensor modes}

The symmetric, traceless and divergence-free tensors, $\gamma_{ij}$ and $\tg_{ij}$, are both gauge invariant and each has two degrees of freedom.
Here, $\gamma_{ij}$ is the gravitational wave coming from the metric fluctuations, while $\tg_{ij}$ is the tensor part of the SU(2) gauge field fluctuations.

 We summarize the above discussion of scalar, vector and tensor modes in the following table. From left to right of the table, we have the fields d.o.f, gauge transformations and finally the number of independent gauge invariant combinations of each part.
\begin{center}
\begin{tabular}{cp{2cm}p{1.7cm}p{1.7cm}p{2cm}p{2cm}p{2.5cm}}
%& & \multicolumn{5}{}{Fields} & \multicolumn{3}{}Equations \\
\hline
&  & $\ \  \ \delta g_{\mu\nu}$ &\begin{footnotesize}
$\ \ \delta\!_{gf}A^a_{\mu}$
\end{footnotesize}  & \begin{footnotesize}
$x^\mu$-gauge
\end{footnotesize} & \begin{footnotesize}
$A^a$-gauge
\end{footnotesize} &\begin{footnotesize}
Gauge-invariant
\end{footnotesize}\\ [2.5ex]
\hline
 &\begin{footnotesize}
Scalar
\end{footnotesize}  & 4 & 4 & $-2$ & $-1$ & 5    \\ [1.1ex] %\hline
& \begin{footnotesize}
Vector
\end{footnotesize} & 2 & 3 & $-1$ & $-1$ & 3     \\[1.1ex] % \hline
& \begin{footnotesize}
Tensor
\end{footnotesize} & 1 & 1 & ~~0  & ~~0 & 2    \\ [1.1ex] %\hline
&\begin{scriptsize}
Total d.o.f
\end{scriptsize} & 10 & 12 & $-4$ & $-3$ & 15 \\[1ex]  \cline{1-7}
\end{tabular}
\vskip 0.1 cm
\textbf{Table II: Perturbed fields and gauge invariant combinations}
\end{center}

In table II, $\delta\!_{gf}A^a_{\mu}$ denotes the genuine gauge field fluctuations, ``$x^\mu$-gauge''
represents the space-time gauge transformations and the ``$A^a$-gauge'' is the internal gauge field transformations.

\section{Computation of the Green's integral of $h^{^{\!s}}_{_{R,L}}$}\label{Green-int}

In this appendix, we determine the explicit form of the inhomogeneous (particular) solution tensor modes, $h^{\!^{s}}_{_{R,L}}$, after horizon crossing. The special part of the gravitational wave is sourced by the gauge field  (Eq. \eqref{h-sigma}) and its wave function can be decomposed as
\bea\label{Integral-I-Ap}
h^{\!^{s}}_{_{R,L}}(\x)= \bigg(\frac{\bar{\rho}_{_{\rm YM}}}{\bar{\rho}}\bigg)^{\!\frac12}\mathcal{G}_{_{R,L}}(\kappa,\mu,\x)h_{_{\rm deS}}(\x),
\eea
where $\x\equiv-k\tau$, $h_{_{\!\rm{deS}}}(\x)$ is the homogeneous wave function solution of \eqref{h-hankel} in de Sitter space
\be
\frac{1}{\sqrt{k}}h_{_{\!\rm{deS}}}(\x)=\frac{1}{\sqrt{2k}}(1+\frac{i}{\x})e^{i\x},
\ee
and $\mathcal{G}_{_{R,L}}(\kappa,\mu,\x)$ is defined by Eqn. \eqref{h-sigma} as
\be
\mathcal{G}_{_{R,L}}(\kappa,\mu,\x)=\frac{e^{i\kappa_{_{R,L}}\!\pi/2}}{\sqrt{(1+\xi_{\psi}^2)/32}}\int^{\infty}_{\x}\frac{G(\x,\x')}{h_{_{\!\rm{deS}}}(\x)\x'}\!\biggl(\!\partial_{\x'}+(\frac{\xi_{\psi}^2}{\x'}\mp\xi_{\psi})\biggl)\!W_{\kappa_{_{R,L}},\mu}(\!-2i\x'\!)d\x'.
\ee
Here $G(\x,\x')$ is the retarded Green's function\footnote{The exact form of the retarded Green's function is \be
 G(\x,\x')=\frac{h(\x)h^*(\x')-h(\x')h^*(\x)}{\textsf{W}\big(h(\x'),h^*(\x')\big)}\Theta(\x'-\x)=\frac{\pi\sqrt{\x\x'}}{2}\big(J_{\nu_T}(\x')Y_{\nu_T}(\x)-J_{\nu_T}(\x)Y_{\nu_T}(\x')\big)\Theta(\x'-\x),\nonumber
\ee
in which $\textsf{W}(h, h^*)$ is the Wronskian of $h$ and $h^*$,
$\textsf{W}(h, h^*)=i$, while 
 $J_{\nu}$ and $Y_{\nu}$ are the first and second kind of Bessel functions. However, the source term $\Pi^T_{_{R}}$ is only important during the tachyonic phase of $\th_{_{R}}$ which is before horizon crossing and hence we can neglect the slow-roll terms in $h$. Using the de Sitter approximation 
 $h(\x)\simeq \frac{1}{\sqrt{2k}}(1+\frac{i}{\x})e^{i\x}$ in the above Green's function, we then obtain \eqref{Green-Ap}.}
\bea
\label{Green-Ap}
G(\x,\x')\simeq\bigg(\frac{\x'-\x}{\x'\x}\cos(\x'-\x)-(1+\frac{1}{\x\x'})\sin(\x'-\x)\bigg)\Theta(\x'-\x),
\eea
where $\Theta(\x-\x')$ is the Heveside's delta function.

Inserting \eqref{Green-Ap} into \eqref{Integral-I-Ap}, the integral at super-horizon scales is given as
\bea\label{Integral-I-App}
\mathcal{G}_{_{_{\sigma}}}(\kappa_{\sigma},\mu)\!\simeq \frac{8e^{i\kappa_{_{R,L}}\!\pi/2}}{\sqrt{(1+\xi_{\psi}^2)}} \!\int\!\frac{1}{\x'}\!\big(\cos\x'-\frac{\sin\x'}{\x'}\big)\bigg(\!\partial_{\x'}W_{\kappa_{\sigma},\mu}+(\frac{\xi^2_{\psi}}{\x'}\mp\xi_{\psi})W_{\kappa_{\sigma},\mu}\!\bigg)d\x'\rvert_{\x'=\x_0},~~~~
\eea
where $\x_0\equiv-k\tau_0$ and $\tau_0$ is the beginning of inflation ($\x_0\gg1$).

The Whittaker functions satisfy the following integral identities
\bea
\int x^n e^{-ix}W_{\mu,\kappa}(-2ix)dx&=&\frac{x^{n+1}\textmd{G}^{2,2}_{2,3}\biggl(-2ix\bigg|\begin{matrix}
  -n, & 1+\kappa & \\
  \frac12-\mu, & \mu+\frac12, & -n-1\\
  \end{matrix}\biggl)}{\Gamma(\frac12-\kappa-\mu)\Gamma(\frac12-\kappa+\mu)},\\
\int x^n e^{ix}W_{\mu,\kappa}(-2ix)dx&=&x^{n+1}\textmd{G}^{2,1}_{2,3}\biggl(-2ix\bigg|\begin{matrix}
  -n, & 1-\kappa & \\
  \frac12-\mu, & \mu+\frac12, & -n-1\\
  \end{matrix}\biggl).
\eea 
Making use of the above identities and doing the integral \eqref{Integral-I-App}, we obtain
\bea\label{Int-I}
&&\mathcal{G}(\kappa,\mu)\!\simeq\! \frac{e^{i\kappa\!\pi/2}}{\sqrt{(1+\xi_{\psi}^2)/2}}\bigg[\! -(i+\xi_{\psi})\big(\frac{\textmd{G}^{2,1}_{2,3}\biggl(-2i\x\bigg|\begin{matrix}
     1, & 1+\kappa & \\
      \frac12-\mu, & \frac12+\mu, & 0\\
   \end{matrix}\bigg)}{\Gamma(\frac12-\kappa-\mu)\Gamma(\frac12-\kappa+\mu)}
  +\textmd{G}^{2,2}_{2,3}\biggl(-2i\x\bigg|\begin{matrix}
  1, & 1-\kappa & \\
 \frac12-\mu, & \frac12+\mu, & 0\\
  \end{matrix}\bigg)\big)\nonumber\\
  &-&\frac{1}{\x}\textmd{G}^{2,2}_{2,3}\biggl(-2i\x\bigg|\begin{matrix}
  2, & -\kappa & \\
 \frac12-\mu, & \frac12+\mu, & 1\\
   \end{matrix}\biggl)+\frac{(1-\kappa-i\xi_{\psi}+\xi^2_{\psi})}{\x}\textmd{G}^{2,2}_{2,3}\biggl(-2i\x\bigg|\begin{matrix}
               2, & 1-\kappa & \\
               \frac12-\mu, & \frac12+\mu, & 1\\
               \end{matrix}\biggl)\nonumber\\
&-&\frac{1}{\x}\frac{\textmd{G}^{2,1}_{2,3}\biggl(-2i\x\bigg|\begin{matrix}
       2, & 2+\kappa & \\
      \frac12-\mu, & \frac12+\mu, & 1\\
        \end{matrix}\bigg)}{\Gamma(-\frac12-\kappa-\mu)\Gamma(-\frac12-\kappa+\mu)}
          -\frac{(1+\kappa-i\xi_{\psi}-\xi^2_{\psi})}{\x}\frac{\textmd{G}^{2,1}_{2,3}\biggl(-2i\x\bigg|\begin{matrix}
           2, & 1+\kappa & \\
          \frac12-\mu, & \frac12+\mu, & 1\\
         \end{matrix}\bigg)}{\Gamma(\frac12-\kappa-\mu)\Gamma(\frac12-\kappa+\mu)}\nonumber\\
&+&\frac{i(\xi^2_{\psi}-\kappa)}{\x^2}\big(\textmd{G}^{2,2}_{2,3}\biggl(-2i\x\bigg|\begin{matrix}
 3, & 1-\kappa & \\
  \frac12-\mu, & \frac12+\mu, & 2\\
  \end{matrix}\bigg)-\frac{\textmd{G}^{2,1}_{2,3}\biggl(-2i\x\bigg|\begin{matrix}
               3, & 1+\kappa & \\
              \frac12-\mu, & \frac12+\mu, & 2\\
             \end{matrix}\bigg)}{\Gamma(\frac12-\kappa-\mu)\Gamma(\frac12-\kappa+\mu)}\big)\nonumber\\
 &-&\frac{i}{\x^2}\textmd{G}^{2,2}_{2,3}\biggl(-2i\x\bigg|\begin{matrix}
  3, & -\kappa & \\
 \frac12-\mu, & \frac12+\mu, & 2\\
  \end{matrix}\bigg)
 +\frac{i}{\x^2}\frac{\textmd{G}^{2,1}_{2,3}\biggl(-2i\x\bigg|\begin{matrix}
     3, & 2+\kappa & \\
    \frac12-\mu, & \frac12+\mu, & 2\\
   \end{matrix}\bigg)}{\Gamma(-\frac12-\kappa-\mu)\Gamma(-\frac12-\kappa+\mu)}\bigg]\rvert_{\x'=\x_0}.        
\eea
The G-function with $Re(p)>0$, $Re(q)>0$ and $p-q\neq0$, has the following asymptotic form for $x\gg1$
\bea
&&\frac{1}{x^{p-1}}\textmd{G}^{2,1}_{2,3}\bigg(-2ix\bigg|\begin{matrix}
 p, & q & \\
 \frac12-\mu, & \frac12+\mu, & p-1\\
 \end{matrix}\bigg)\simeq\frac{i(-2i)^p}{2}\Gamma(\frac32-p-\mu)\Gamma(\frac32-p+\mu)\Gamma(p-q),\nonumber\\
&+&\frac{i(-2i)^q}{2(q-p)}\Gamma(\frac32-q-\mu)\Gamma(\frac32-q+\mu)x^{q-p}.\nonumber\\
&&\frac{1}{x^{p-1}}\textmd{G}^{2,2}_{2,3}\bigg(-2ix\bigg|\begin{matrix}
  p, & q & \\
 \frac12-\mu, & \frac12+\mu, & p-1\\
  \end{matrix}\bigg)\simeq\frac{i(-2i)^p}{2}\frac{\Gamma(\frac32-p-\mu)\Gamma(\frac32-p+\mu)}{\Gamma(1-p+q)}.
\eea
Upon using the above relations in \eqref{Int-I}, we obtain
\bea\label{Int-III}
&&\mathcal{G}(\kappa,\mu)\!\simeq\! \bigg[-\frac{2(\kappa-\xi^2_{\psi})\Gamma(2-\kappa)\Gamma(-\frac32-\mu)\Gamma(-\frac32+\mu)}{\Gamma(\frac12-\kappa-\mu)\Gamma(\frac12-\kappa+\mu)}-\frac{(1-\kappa+\xi^2_{\psi}-i\xi_{\psi})\Gamma(-\frac12-\mu)\Gamma(-\frac12+\mu)}{\Gamma(-\kappa)}~~~~~\nonumber\\ 
&&-2i\frac{\Gamma(-\kappa)\Gamma(-\frac12-\mu)\Gamma(-\frac12+\mu)-2\Gamma(1-\kappa)\Gamma(-\frac32-\mu)\Gamma(-\frac32+\mu)}{\Gamma(-\frac12-\kappa-\mu)\Gamma(-\frac12-\kappa+\mu)}+\frac{(i\xi_{\psi}-1)\Gamma(\frac12-\mu)\Gamma(\frac12+\mu)}{2\Gamma(1-\kappa)}\nonumber\\
&+&\frac{(i\xi_{\psi}-1)\Gamma(-\kappa)\Gamma(\frac12-\mu)\Gamma(\frac12+\mu)+2(1+\kappa-i\xi_{\psi}-\xi^2_{\psi})\Gamma(1-\kappa)\Gamma(-\frac12-\mu)\Gamma(-\frac12+\mu)}{2\Gamma(\frac12-\kappa-\mu)\Gamma(\frac12-\kappa+\mu)}\nonumber\\
&+&\frac{\Gamma(-\frac12-\mu)\Gamma(-\frac12+\mu)+2(\kappa-\xi^2_{\psi})\Gamma(-\frac32-\mu)\Gamma(-\frac32+\mu)}{\Gamma(-1-\kappa)}+2\frac{\Gamma(-\frac32-\mu)\Gamma(-\frac32+\mu)}{\Gamma(-2-\kappa)}\bigg]\frac{4e^{i\kappa\pi/2}}{\sqrt{(1+\xi_{\psi}^2)}}
.\nonumber
\eea
Using the slow-roll relation \eqref{xi-eq}, we can read $\mu$ and $\kappa$ in terms of $\xi_{\psi}$ as
\be\label{kappa-mu}
\kappa_{_{R,L}}=\mp i\bigg(\frac{1+2\xi^2_{\psi}}{\xi_{\psi}}\bigg) \quad \textmd{and} \quad \mu^2=\frac14-2(1+\xi^2_{\psi}),
\ee
which implies that $\mathcal{G}(\kappa,\mu)$ is simply a function of $\xi_{\psi}$.
Recalling the functional equation $\Gamma(x+1)=x\Gamma(x)$ for $Re(x)\geq0$, we can write $\mathcal{G}(\kappa_{\sigma},\mu)$ as
\bea\label{Int-IV}
&&\mathcal{G}_{\sigma}(\xi_{\psi})\!\simeq\!\bigg[
-\bigg(\frac{(i\xi_{\psi}+1)}{2}+\frac{\kappa_{\sigma}(i\xi_{\psi}+\xi^2_{\psi}+2)}{(\frac14-\mu^2)}
+\frac{2\kappa_{\sigma}(\kappa_{\sigma}-1)(2+\xi^2_{\psi})}{(\frac14-\mu^2)(\frac94-\mu^2)}
\bigg)\frac{\Gamma(-\kappa_{\sigma})\Gamma(\frac12-\mu)\Gamma(\frac12+\mu)}{\Gamma(\frac12-\kappa_{\sigma}-\mu)\Gamma(\frac12-\kappa_{\sigma}+\mu)}\nonumber\\
&+&\bigg(\frac{(2+\xi^2_{\psi}-i\xi_{\psi})}{(\frac14-\mu^2)}-\frac{(1-i\xi_{\psi})}{2\kappa_{\sigma}}-\frac{2(1+\kappa_{\sigma})(2+\xi^2_{\psi})}{(\frac94-\mu^2)(\frac14-\mu^2)}\bigg)\frac{\Gamma(\frac12-\mu)\Gamma(\frac12+\mu)}{\Gamma(-\kappa_{\sigma})}\bigg]\frac{4e^{i\kappa_{\sigma}\pi/2}}{\sqrt{(1+\xi_{\psi}^2)}}.
\eea
Recalling that $i\kappa_{_{R,L}}\in\mathbb R$
\be
i\kappa_{_{R,L}}=\pm\textmd{sign}(\psi)\bigg(\frac{1+2\xi^2_{\psi}}{|\xi_{\psi}|}\bigg),
\ee
equation \eqref{Int-IV} implies that $\mathcal{G}(\xi_{\psi})$ is subleading for the polarization state with $i\kappa_{\sigma}<0$. As a result, we only need to determine $\mathcal{G}_{\sigma}(\xi_{\psi})$ for the polarization with $i\kappa_{\sigma}>0$, $\mathcal{G}_{+}(\xi_{\psi})$. Using the slow-roll relations $\frac94-\mu^2\simeq(2+\xi_{\psi}^2)$ and $i\kappa\!_{_{+}}\simeq\frac{(1+2\xi_{\psi}^2)}{|\xi_{\psi}|}$, we can mostly simplify $\mathcal{G}_{+}$ as
\bea\label{Int-IIV}
\mathcal{G}\!_{_{+}}(\xi_{\psi})\!&\simeq&\!e^{\frac{i\kappa\!_{_{+}}\pi}{2}}\frac{2\sqrt{(1+\xi^2_{\psi})}}{\xi_{\psi}^2} \bigg(
\frac{(i\xi_{\psi}+1)\Gamma(-\kappa\!_{_{+}})}{\Gamma(\frac12-\kappa\!_{_{+}}-\mu)\Gamma(\frac12-\kappa\!_{_{+}}+\mu)}+\frac{(i\xi_{\psi}-1)}{\Gamma(1-\kappa\!_{_{+}})}\bigg)\Gamma(\frac12-\mu)\Gamma(\frac12+\mu).~~~~~~~~~~
\eea
The functions $\mathcal{G}_{\pm}(\xi_{\psi})$ are plotted with respect to $\xi_{\psi}$ in figure \ref{IR-Gamma}.

\end{document}